\documentclass[journal]{IEEEtran}
\usepackage[applemac]{inputenc}
\usepackage[english]{babel}

\usepackage{amsmath}
\usepackage{amssymb}
\usepackage{dsfont}
\usepackage{amsthm}
\usepackage{thmtools}
\usepackage{cases}
\usepackage{balance}

\newtheorem{lemma}{Lemma}

\newtheorem{theorem}{Theorem}
\newtheorem{corollary}{Corollary}
\newtheorem{remark}{Remark}

\newtheoremstyle{mydef}
	{3pt}		
	{3pt}		
	{}		
	{}		
	{\itshape}	
	{:}		
	{.5em}	
	{}		

\theoremstyle{mydef}

\usepackage{cite} 

\usepackage[nolist, nohyperlinks]{acronym}

\acrodefplural{t.u.}[t.u.]{time units}

\ifCLASSOPTIONcompsoc
	\usepackage[caption=false,font=normalsize,labelfont=sf,textfont=sf]{subfig}
\else
	\usepackage[caption=false,font=footnotesize]{subfig}
\fi

\usepackage{tikz}
\usetikzlibrary{calc,
shapes,
positioning,
snakes,
patterns}

\definecolor{green}{RGB}{34,150,34}

\renewcommand{\P}{\boldsymbol{P}}
\newcommand{\G}{\boldsymbol{G}}

\newcommand{\cc}{\mathsf{c}}
\newcommand{\dd}{\mathsf{d}}
\newcommand{\x}{\boldsymbol{x}}
\newcommand{\X}{\boldsymbol{X}}
\newcommand{\q}{\boldsymbol{q}}
\newcommand{\y}{\boldsymbol{y}}

\newcommand{\gBS}{\gamma_\text{\textnormal{BS}}}
\newcommand{\gD}{\gamma_\text{\textnormal{D2D}}}

\newcommand{\mrBS}{m_{\text{\textnormal{r}}}^\text{\textnormal{BS}}}
\newcommand{\mrD}{m_{\text{\textnormal{r}}}^\text{\textnormal{D2D}}}
\newcommand{\mrgD}{m_\text{\textnormal{r,g}}^\text{\textnormal{D2D}}}
\newcommand{\mrlD}{m_\text{\textnormal{r,l}}^\text{\textnormal{D2D}}}

\newcommand{\rhoBS}{\rho_\text{\textnormal{BS}}}
\newcommand{\rhoD}{\rho_\text{\textnormal{D2D}}}

\newcommand{\aLRC}{\alpha_\textnormal{LRC}}
\newcommand{\aMBR}{\alpha_\text{MBR}}
\newcommand{\aMDS}{\alpha_\text{MDS}}
\newcommand{\aMSR}{\alpha_\text{MSR}}
\newcommand{\arep}{\alpha_\text{rep}}
\newcommand{\bLRC}{\beta_\text{LRC}}
\newcommand{\bMBR}{\beta_\text{MBR}}
\newcommand{\bMDS}{\beta_\text{MDS}}
\newcommand{\bMSR}{\beta_\text{MSR}}

\newcommand{\dr}{\Delta}
\newcommand{\drmax}{\Delta_\text{max}}
\newcommand{\dropt}{\Delta_\text{opt}}

\newcommand{\C}{\bar{C}}
\newcommand{\Cd}{\bar{C}_\text{\textnormal{d}}}
\newcommand{\Cr}{\bar{C}_\text{\textnormal{r}}}

\newcommand{\CdBS}{\bar{C}_\text{\textnormal{d}}^\text{\textnormal{BS}}}
\newcommand{\CrBS}{\bar{C}_\text{\textnormal{r}}^\text{\textnormal{BS}}}

\newcommand{\Ch}{\bar{C}^\text{\textnormal{hybrid}}}
\newcommand{\Cdh}{\bar{C}_\text{\textnormal{d}}^\text{\textnormal{hybrid}}}
\newcommand{\Crh}{\bar{C}_\text{\textnormal{r}}^\text{\textnormal{hybrid}}}

\renewcommand{\Pr}{\mathbb{P}}

\newcommand{\PrBSd}{p_\text{\textnormal{BS}}}
\newcommand{\PrDd}{p_\text{\textnormal{D2D}}}

\newcommand{\Ta}{T_\text{a}}
\newcommand{\Tl}{T_\text{l}}
\newcommand{\Tr}{T_\text{r}}

\newcommand{\Wlt}{\tilde W_\ell}
\newcommand{\Winf}{\tilde W_{\infty}}

\IEEEoverridecommandlockouts

\title{Distributed Storage in Mobile Wireless Networks with Device-to-Device Communication}

\begin{document}


\begin{acronym}[OFDM]
	\acro{BS}{base station} 
	\acro{cdf}{cumulative distribution function}
	\acro{CDN}{content delivery network}
	\acro{c.u.}{cost units}
	\acro{D2D}{\textit{device-to-device}} 
	\acro{DS}{distributed storage} 
	\acro{ECC}{erasure correcting code}
	\acro{i.i.d.}{independent, identically distributed} 
	\acro{LRC}{locally repairable code} 
	\acro{MBR}{minimum bandwidth regenerating} 
	\acro{MDS}{maximum distance separable} 
	\acro{MIMO}{multiple input multiple output}
	\acro{MSR}{minimum storage regenerating} 
	\acro{OFDM}{orthogonal frequency division multiplexing}
	\acro{P2P}{peer-to-peer}
	\acro{pdf}{probability density function} 
	\acro{pmf}{probability mass function} 
	\acro{RV}{random variable}
	\acro{t.u.}{time unit}
\end{acronym}

\author{Jesper Pedersen, Alexandre Graell i Amat,~\IEEEmembership{Senior Member,~IEEE},\\ Iryna Andriyanova,~\IEEEmembership{Member,~IEEE}, and Fredrik Br\"annstr\"om,~\IEEEmembership{Member,~IEEE}
\thanks{This paper was presented in part at the IEEE Information Theory Workshop, Jeju Island, Korea, October 2015.}
\thanks{This work was partially funded by the Swedish Research Council under grants 2011-5961 and 2011-5950, and by the European Research Council under Grant No. 258418 (COOPNET).}
\thanks{J. Pedersen, A. Graell i Amat, and F. Br\"annstr\"om are with the Department of Signals and Systems, Chalmers University of Technology, SE-41296 Gothenburg, Sweden (e-mail: \{jesper.pedersen, alexandre.graell, fredrik.brannstrom\}@chalmers.se).}
\thanks{I. Andriyanova is with the ETIS-UMR8051 group, ENSEA/University of Cergy-Pontoise/CNRS, 95015 Cergy, France (e-mail: iryna.andriyanova@ensea.fr).}
\vspace{-3ex}}

\maketitle

\begin{abstract}
We consider the use of distributed storage (DS) to reduce the communication cost of content delivery in wireless networks. Content is stored (\emph{cached}) in a number of mobile devices using an erasure correcting code. Users retrieve content from other devices using device-to-device communication or from the base station (BS), at the expense of higher communication cost. We address the repair problem when a device storing data leaves the cell. We introduce a repair scheduling where repair is performed periodically and derive analytical expressions for the overall communication cost of content download and data repair as a function of the repair interval. The derived expressions are then used to evaluate the communication cost entailed by DS using several erasure correcting codes. Our results show that DS can reduce the communication cost with respect to the case where content is downloaded only from the BS, provided that repairs are performed frequently enough. If devices storing content arrive to the cell, the communication cost using DS is further reduced and, for large enough arrival rate, it is always beneficial. Interestingly, we show that MDS codes, which do not perform well for classical DS, can yield a low overall communication cost in wireless DS.

\end{abstract}



\section{Introduction}
\IEEEPARstart{I}{t is} predicted that the global mobile data traffic will exceed 30 exabytes per month by 2020, nearly a tenfold increase compared to the traffic in 2015 \cite{Cisco2015}. This dramatic increase threatens to completely congest the already burdened wireless networks. One popular approach to reduce peak traffic is to store popular content closer to the end users, a technique known as \textit{caching}. 
The idea is to deploy a number of access points (called helpers) with large storage capacity, but low-rate wireless backhaul, and store data across them \cite{Shanmugam2013, Bioglio2015}. Users can then download content from the helpers, resulting in a higher throughput per user. In \cite{Golrezaei2014} it was suggested to store content directly in the mobile devices, taking advantage of the high storage capacity of modern smart phones and tablets. The requested content can then be directly retrieved from neighbouring mobile devices, using \ac{D2D} \textit{communication}. This allows for a more efficient content delivery at no additional infrastructure cost. 
Caching in the mobile devices to alleviate the wireless bottleneck has attracted a significant interest in the research community in the recent years  \cite{Maddah-Ali2014, Golrezaei2014:scaling, Yang2015, Ji2016}. In all these works, simple content caching and/or replication (i.e., a number of copies of a content are stored in the network) is considered. Additionally, the use of \ac{MDS} codes to facilitate decentralized random caching was investigated in \cite{Ji2016}.

A relevant problem in \ac{D2D}-assisted mobile caching networks is the repairing of the lost data when a storage device is unavailable, e.g., when a storage device fails or leaves the network. Repairing of the lost data was considered in \cite{Paakkonen2013}, where the communication cost incurred by data download and repair was analyzed for a caching scheme where data is stored in the mobile devices using replication and regenerating codes \cite{Dimakis2010}. A strong assumption in \cite{Paakkonen2013} is that the repair of the lost content is performed instantaneously. As a result, content can always be downloaded from the mobile devices. Under the assumption of instantaneous repair, the caching strategy that minimizes the overall communication cost is $2$-replication.

In this paper, we consider content caching in a wireless network scenario using erasure correcting codes. When using erasure correcting codes to cache content, caching bears strong ties with the concept of \ac{DS} for reliable data storage. Indeed, the set of mobile devices storing content can be seen as a distributed storage network. The fundamental difference with respect to \ac{DS} for reliable data storage is that data download can be done not only from the storage nodes, but the \ac{BS} can also assist to deliver the data. Therefore, the strict guarantees on fault tolerance can be relaxed, which brings new and interesting degrees of freedom with respect to erasure-correcting coding for \ac{DS} for reliable data storage. 
Here, to avoid confusion with standard (uncoded) caching, we will use the term {\em wireless distributed storage}, highlighting the resemblance with \ac{DS} using erasure correcting codes for reliable data storage in, e.g., data centers. Similar to the scenario in \cite{Paakkonen2013}, we consider a cellular system where mobile devices roam in and out of a cell according to a Poisson random process and request content at random times. The cell is served by a \ac{BS}, which always has access to the content. Content is also stored across a limited number of mobile devices using an erasure correcting code. 
Our main focus is on the repair problem when a device that stores data leaves the network. In particular, we introduce a more realistic repair scheduling than the one in \cite{Paakkonen2013} where lost content is repaired (from storage devices using \ac{D2D} communication or from the \ac{BS}) at periodic times. 

We derive analytical, closed-form expressions for the overall communication cost of content download and data repair as a function of the repair interval. The derived expressions are general and can be used to analyze the overall communication cost incurred by any erasure correcting code for \ac{DS}. As an example of the application of the proposed framework, we analyze the overall communication cost incurred by \ac{MDS} codes, regenerating codes \cite{Dimakis2010}, and \acp{LRC} \cite{Papailiopoulos2012}. 
We show that wireless \ac{DS} can reduce the overall communication cost as compared to the basic scenario where content is only downloaded from the \ac{BS}. However, this is provided that repairs can be performed frequently enough. Moreover, in the case when  nodes storing content arrive to the cell, the communication cost using \ac{DS} is further reduced and, for large enough arrival rate, it is always beneficial as compared to \ac{BS} download. The repair interval that minimizes the overall communication cost depends on the network parameters and the underlying erasure correcting code. We show that, in general, instantaneous repair is not optimal. The derived expressions can also be used to find, for a given repair interval, the erasure correcting code yielding the lowest overall communication cost.


Non-instantaneous repairs, the so-called ``lazy'' repairs, have already been proposed for \ac{DS} in data centers \cite{Giroire2010,Silberstein2014} to reduce the amount of data that has to be transmitted within the storage network during
the repair process, known as the {\em repair bandwidth}. However, contrary to \cite{Giroire2010,Silberstein2014}, in the wireless scenario considered here the non-instantaneous repairs impact both data repair and  download. We show that, somewhat interestingly, erasure correcting codes achieving a low repair bandwidth do not always perform well in a wireless \ac{DS} setting. On the other hand, MDS codes, which entail a high repair bandwidth, can yield a low overall communication cost for some repair intervals.



\textit{Notation:} The \ac{pdf} of a random variable $X$ is denoted by $f_X(\cdot)$. Expectation and probability are denoted by $\mathbb{E}[\hspace{.25ex}\cdot\hspace{.25ex}]$ and $\mathbb{P}(\cdot)$, respectively. We use bold lowercase letters $\boldsymbol{x}$ to denote vectors and bold uppercase letters $\boldsymbol{X}$ for matrices.

\section{System Model}
\label{sec:model}

\begin{figure}[!t]
	\centering
	\includegraphics[width=1\columnwidth]{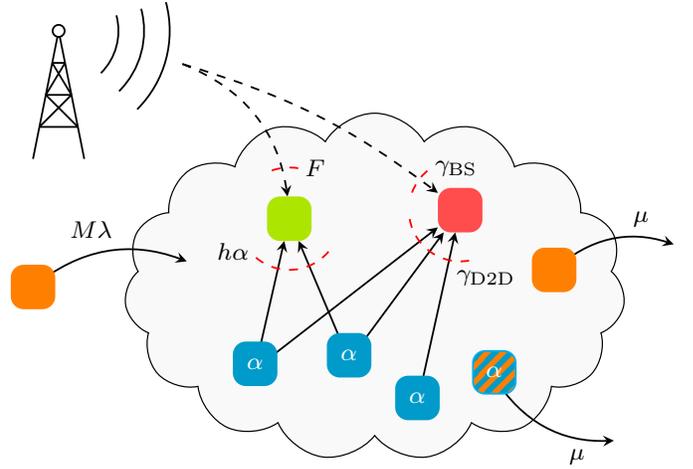}
	\vspace{-3ex}
	\caption{A wireless network with data storage in the mobile devices (nodes). A new node arrives to the network at rate $M \lambda$. The departure rate per node is $\mu$. Blue nodes store exactly $\alpha$ bits each. The green node requests the file and downloads it from the storage nodes (solid arrows), or from the \ac{BS} (dashed arrow). The repair onto a node (in red) is carried out by transmitting $\gD$ bits from storage nodes (solid arrows) or $\gBS$ bits from the \ac{BS} (dashed arrow).
	}
	\label{fig:sys}
	\vspace{-3ex}
\end{figure}

We consider a single cell in a cellular network, served by a \ac{BS}, where mobile devices (referred to as nodes) arrive and depart according to a Poisson random process. The initial number of nodes in the network is $M$. Nodes wish to download content from the network. For simplicity, we assume that there is a single object (file), of size $F$ bits, stored at the \ac{BS}. We further assume that nodes can store data and communicate between them using \ac{D2D} communication. The considered scenario is depicted in Fig.~\ref{fig:sys}.

{\it Arrival-departure model.} 
Nodes arrive according to a Poisson process with exponential  \ac{i.i.d.} random inter-arrival times $\Ta$ with \ac{pdf}
\begin{equation}\label{eq:Tapdf}
	f_{\Ta}(t)=M\lambda e^{-M\lambda t},\quad \lambda\ge0,~t\geq0,
\end{equation}
where $M\lambda$ is the expected arrival rate of a node and $t$ is time, measured in \acp{t.u.}.

The nodes stay in the cell for an \ac{i.i.d.} exponential random lifetime $\Tl$ with pdf
\begin{equation}\label{eq:Tlpdf}
f_{\Tl}(t)=\mu e^{-\mu t},\quad \mu\ge0,~t\geq0,
\end{equation}
where $\mu$ is the expected departure rate of a node. The number of nodes in the cell can be described by an $\mathsf{M}/\mathsf{M}/\infty$ queuing model where the probability that there are $i$ nodes in the cell is \cite{Miller2004}
\begin{equation}
	\label{eq:pi}
	\pi(i)=\frac{(M\lambda/\mu)^i}{i!}e^{-(M\lambda/\mu)}.
\end{equation}
For simplicity, we assume that $\mu=\lambda$, i.e., the flow in and out from the cell is the same and the expected number of nodes in the cell stays constant (equal to $M$).

{\it Data storage.}
The file is partitioned into $k$ packets, called symbols, of size $\frac{F}{k}$ bits and is encoded into $n$ coded symbols, $n\ge 2$, using an $(n,k)$ erasure correcting code of rate $R=k/n<1$. The encoded data is stored in $m$ nodes, $2\le m \le n$, referred to as \emph{storage nodes}. Note that $m\le n$ implies that a storage node may store multiple coded symbols. For some of the considered erasure correcting codes, this is the case (see Section~\ref{sec:codes}). To simplify the analysis in Sections~\ref{sec:analysis} and \ref{sec:hybrid}, we set $m\ll M$. This guarantees that the probability that the number of nodes in the cell is smaller than $m$ is negligibly small, i.e.,
\begin{equation}
	\label{eq:empty}
	\sum_{i=0}^{m-1}\pi(i)\ll 1,
\end{equation}
using \eqref{eq:pi}. For example, for $m\le10$ and $M=30$, \eqref{eq:empty} is less than $7.2\cdot10^{-6}$. Therefore, with high probability the file can be stored in the cell. In the results section we show that this simplification has negligible impact and that the analytical expressions match closely with the simulation results.

Each storage node stores exactly $\alpha$ bits, i.e., we consider a symmetric allocation \cite{Leong2012}. Hence\footnote{Without loss of generality, we assume $\alpha\in\mathbb{N}$.},
\begin{align}
	\label{eq:alpha}
	\alpha =\frac{1}{m}\cdot\frac{F}{R} \ge \frac{F}{k}.
\end{align}

{\it Incoming process.} Nodes arriving to the cell may bring cached content. The expected arrival rate of nodes storing content is $m\lambda_\text{c}$, $\lambda_\text{c}\le\mu$. We also assume that the expected arrival rate of nodes not carrying content is $M\lambda-m\lambda_\text{c}$, so that the expected arrival rate of a node (with or without content) is $M\lambda$ and the expected number of nodes in the cell is $M$ (see above). The incoming process is discussed in more detail in Section~\ref{sec:Incoming}.


{\it Data delivery.}
Nodes request the file at random times with \ac{i.i.d.} random inter-request time $\Tr$ with pdf
\begin{equation}\label{eq:Trpdf}
	f_{\Tr}(t)=\omega e^{-\omega t},\quad \omega\geq0,~t\geq0,
\end{equation}
where $\omega$ is the expected request rate per node. Whenever possible, the file is downloaded from the storage nodes using \ac{D2D} communication, referred to as \ac{D2D} download. In particular, we assume that data can be downloaded from any subset of $h$ storage nodes, $1\le h< m$, which we will refer to as the \emph{download locality}. In other words, \ac{D2D} download is possible if $h$ or more storage nodes remain in the cell. In this case, the amount of downloaded data is $h\alpha \ge  F$ bits.\footnote{To simplify the analysis in Sections~\ref{sec:analysis} and~\ref{sec:hybrid}, we assume that the download bandwidth is the same irrespective of whether the request comes from a storage node itself or not, i.e., users do not have access to their own stored data. This is a reasonable approximation if $m \ll M$. Furthermore, this may be a practical assumption. Due to concerns about security in systems that allow for \ac{D2D} connectivity, it has been proposed to isolate part of the memory in the mobile devices to be used only for \ac{DS}, so that devices cannot have access to their own cached data \cite{Golrezaei2013}.} 
In the case where there are less than $h$ storage nodes in the cell, the file is downloaded from the \ac{BS}, which we refer to as \ac{BS} download. In this case, $F$ bits are downloaded.

{\it Communication cost.}
We assume that transmission from the \ac{BS} and from a storage node (in \ac{D2D} communication) have different costs. We denote by $\rhoBS$ and $\rhoD$ the cost (in \ac{c.u.} per bit, [c.u./bit]) of transmitting one bit from the \ac{BS} and from a storage node, respectively. Therefore, the cost of downloading a file from the \ac{BS} and the storage nodes is $\rhoBS F$ and $\rhoD h\alpha$, respectively. Furthermore, we define $\rho\triangleq\rhoBS/\rhoD>0$, where $\rho>1$ corresponds to a high traffic load in the \ac{BS}-to-device link and $\rho<1$ reflects a scenario where the battery of the devices is the main constraint.


\subsection {Repair Process}
\label{sec:RepairProcess}

When a storage node leaves the cell, its stored data is lost (see blue node with orange stripes in Fig.~\ref{fig:sys}). Therefore, 
another node needs to be populated with data to maintain the initial state of reliability of the \ac{DS} network, i.e., $m$ storage nodes. The restore (repair) of the lost data onto another node, chosen uniformly at random from all nodes in the cell that do not store any content, will be referred to as the repair process. We introduce a scheduled repair scheme where the repair process is run periodically. We denote the interval between two repairs by $\dr$ (in \acp{t.u.}), $\dr \ge 0$. Note that $\dr=0$ corresponds to the case of instantaneous repair, considered in \cite{Paakkonen2013}.

Similar to the download, repair can be accomplished from the storage nodes (\ac{D2D} repair) or from the \ac{BS} (\ac{BS} repair), with cost per bit $\rhoD$ and $\rhoBS$, respectively. The amount of data (in bits) that needs to be retrieved from the network to repair a single failed node is referred to as the \emph{repair bandwidth}, denoted by $\gamma$. For simplicity, we assume that each repair is handled independently of the others. In particular, we assume that \ac{D2D} repair can be performed from any subset of $r$ storage nodes, $1\le r<m$, by retrieving $\beta\le\alpha$ bits from each node. In other words, \ac{D2D} repair is possible if there are at least $r$ storage nodes in the cell at the moment of repair. In this case, $\gD=r\beta\ge\alpha$, and the corresponding communication cost is $\rhoD\gD$. Parameter $r$ is usually referred to as the \emph{repair locality} in the \ac{DS} literature. If there are less than $r$ storage nodes in the cell at the moment of repair, then the repair is carried out by the \ac{BS}. In this case, $\gBS=\alpha$, with communication cost $\rhoBS\gBS$. Note that $\gD/\gBS\ge1$. For both repair and download, we assume error-free transmission.

Parameters $m$, $h$, $r$, $\alpha$ and $\beta$, and subsequently $\gD$ and $\gBS$, depend on the erasure correcting code used for storage. Since $m$, $h$ and $r$ are very important parameters, an erasure correcting code in \ac{DS} is typically defined with the triple $[m,h,r]$. This will be further explained in Section~\ref{sec:codes}.

\section{Repair and Download Cost}
\label{sec:analysis}

In this section, we derive analytical expressions for the repair and download cost, and subsequently for the overall communication cost, as a function of the repair interval $\dr$. For analysis purposes, we initially disregard the incoming process, i.e., set $\lambda_\text{c}=0$. The case $\lambda_\text{c}>0$ is then addressed in Section~\ref{sec:Incoming} building upon the results in this section. We denote by $\Cr$ the average communication cost of repairing lost data, and refer to it as the repair cost. Also, we denote by $\Cd$ the average communication cost of downloading the file, and refer to it as the download cost. The (average) overall communication cost is denoted by $\C$, where $\C\triangleq\Cr+\Cd$. The costs are defined in cost units per bit and time unit, [c.u./(bit$\times$t.u.)]. 

For later use, we denote by $b_i(m,p)$ the \ac{pmf} of the binomial distribution with parameters $m$ and $p$,
\begin{equation}
	\label{eq:binom}
	b_i(m,p) \triangleq \binom{m}{i}p^i(1-p)^{m-i},\quad 0\le i\le m.
\end{equation}

\subsection{Repair Cost}

The repair cost $\Cr$ has two contributions, corresponding to the cases of \ac{BS} repair and \ac{D2D} repair. Denote by $\mrD$ and $\mrBS$ the average number of nodes repaired from the storage nodes and from the \ac{BS}, respectively, in one repair interval. Then, $\Cr$ (in [c.u./(bit$\times$t.u.)]) is given by 
\begin{align}
\label{eq:Cr}
\Cr  = \frac{1}{F\dr}\left(\rhoBS \gBS \mrBS +\rhoD\gD\mrD\right),
\end{align}
where $\rhoBS \gBS$ and $\rhoD\gD$ (in c.u.) are the cost of repairing a single storage node from the \ac{BS} and from storage nodes, respectively (see Section~\ref{sec:RepairProcess}), and we normalize by $F$ such that $\Cr$ does not depend on the file size. 

The repair cost, $\Cr$, is given in the following theorem.
\begin{theorem}
\label{thm:Cr}
	Consider the DS network in Section~\ref{sec:model} with departure rate $\mu$, communication costs $\rhoBS$ and $\rhoD$, BS repair bandwidth $\gBS$, file size $F$, repair interval $\Delta$, and probability $p$ that a node has not left the network during a time $\Delta$. Furthermore, consider the use of an $[m,h,r]$ erasure correcting code with D2D repair bandwidth $\gD$. The repair cost is given by
\begin{align}
\label{eq:CrT}
	\Cr=\frac{1}{F\dr}&\left(\rhoBS \gBS \sum_{i=0}^{r-1}(m-i) b_i(m, p)\right.\nonumber\\
		&\left.+\rhoD\gD \sum_{i=r}^m(m-i) b_i(m, p)\right).
\end{align}
\end{theorem}
\begin{IEEEproof}
	As the inter-departure times are exponentially distributed, the probability that a storage node has not left the network during a time $\dr$ and is available for repair is
\begin{equation*}
	p=\Pr(\Tl>\dr)=e^{-\mu\dr}.
\end{equation*}
Hence, the probability that $i$ storage nodes are available for repair is $b_i(m,p)$. If $i$ storage nodes remain in the cell, then $m-i$ repairs need to be performed. \ac{D2D} repair is performed if $i \ge r$, and \ac{BS} repair is performed otherwise. Therefore,
\begin{align*}
		\mrD  = \sum_{i=r}^{m}(m-i) b_i(m, p),\quad
		\mrBS  = \sum_{i=0}^{r-1}(m-i) b_i(m, p).
	\end{align*}
Using these expressions in \eqref{eq:Cr}, we obtain (\ref{eq:CrT}).
\end{IEEEproof}
\begin{remark}
	\label{rem:repair}
	We see from \eqref{eq:Cr} that if $\rhoBS\gBS<\rhoD\gD$, i.e., $\rho<\frac{\gD}{\gBS}$, D2D repair should never be performed, as repairing always from the BS yields a lower repair cost. In this case the repair cost would be
	\begin{equation*}
		\CrBS = \frac{1}{F\dr}\rhoBS\gBS m(1-e^{-\mu\dr}).
	\end{equation*}
\end{remark}

\subsection{Download Cost}

Similar to $\Cr$, the download cost $\Cd$ has two contributions, corresponding to the case where content is downloaded from the \ac{BS} and from the storage nodes. Denote by $\PrBSd$ and $\PrDd$ the probability that, for a request, the file is downloaded from the \ac{BS} and from the storage nodes, respectively. Then, $\Cd$ can be written as
\begin{align}
\label{eq:Cda}
	\Cd  = \frac{M\omega}{F}&\big(\rhoBS F \PrBSd
		 +\rhoD h\alpha\PrDd\big),
\end{align}
where $\rhoBS F$ and $\rhoD h\alpha$ are the cost of downloading the file from the \ac{BS} and from the storage nodes, respectively (see Section~\ref{sec:model}), and $M \omega$ is the overall request rate per \ac{t.u.}. Again, we normalize by $F$ so that the cost does not depend on the file size. The download cost is given in the following theorem.
\begin{theorem}
\label{thm:Cd} 
Consider the DS network in Section~\ref{sec:model} with expected number of nodes in the cell $M$, departure rate $\mu$, request rate $\omega$, communication costs $\rhoBS$ and $\rhoD$, file size $F$, and repair interval $\Delta$. Furthermore, consider the use of an $[m,h,r]$ erasure correcting code that stores $\alpha$ bits per node. Let $\mu_i=i\mu$ for $i=h,\ldots,m$, and $p_i = e^{-\mu_i\dr}$. The download cost is given by
	\begin{align}
		\label{eq:Cd}
		\Cd 
		&= M\omega \hspace{-.4ex} \left( \hspace{-.4ex} \rhoBS \hspace{-.3ex} + \hspace{-.3ex} \Bigg( \hspace{-.3ex} \rhoD\frac{h\alpha}{F} \hspace{-.3ex} - \hspace{-.3ex} \rhoBS \right) \hspace{-.4ex} \frac{1}{\dr}\sum_{i=h}^{m}\frac{1-p_i}{\mu_i}\prod_{\substack{j=h\\j\neq i}}^{m}\frac{j}{j-i}\Bigg).
	\end{align}
\end{theorem}

\begin{figure}[t!]
	\centering
	\includegraphics[width=1\columnwidth]{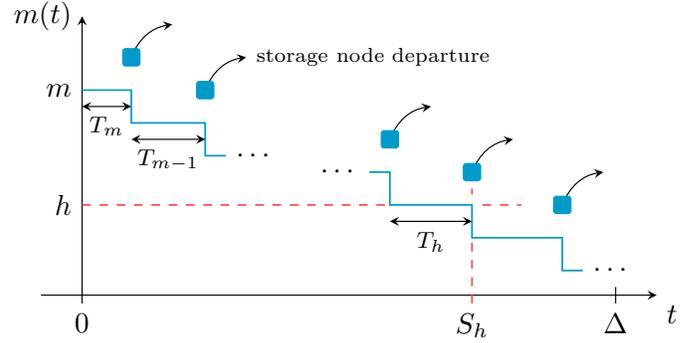}
	\vspace{-4ex}
	\caption{Number of available storage nodes within the repair interval $\dr$. At $t=0$, there are $m$ storage nodes available. $S_h$ is the time after which less than $h$ storage nodes are available, hence \ac{D2D} download is no longer possible.}
	\label{fig:stop}
	\vspace{-2ex}
\end{figure}
The proof is given in Appendix~\ref{prf:Cd}. Here, for ease of understanding, we give an outline of the proof. Since $\PrDd+\PrBSd=1$, it follows from \eqref{eq:Cda} that to derive $\Cd$ is sufficient to derive $\PrDd$. Let $m(t)$ be the number of storage nodes alive in the cell within a repair interval, i.e., for $t\in[0,\Delta)$, with $m(0)=m$.  It is important to observe that $m(t)$ is described by a Poisson death process \cite{Miller2004}, since storage nodes may leave the cell, and no repair is attempted before a time $\Delta$. This random process is illustrated in Fig.~\ref{fig:stop}. At some point, too many storage nodes have left the network, such that the number of available storage nodes goes below $h$ and \ac{D2D} download is no longer possible. Denote the (random) time this occurs by $S_h$, i.e., $m(t)<h\;\forall~t\ge S_h$, $t\in [0, \dr)$ (see Fig.~\ref{fig:stop}). Denote by $\Wlt$ the arrival time of the $\ell$th file request within a repair interval, $t\in [0, \dr)$. The probability $\PrDd$ can then be derived in two steps.
\begin{enumerate}
	\item Find the pdf of the arrival time of the file requests within a repair interval $\Delta$, $\Wlt$.
	\item Find the probability that a request arrives before $S_h$, $\PrDd=\Pr(\Wlt<S_h)$ (i.e., \ac{D2D} download is possible).
\end{enumerate}

%
%

\begin{remark}
	\label{rem:download}
	If $\rhoBS F<\rhoD h\alpha$, i.e., $\rho<\frac{h\alpha}{F}$, performing BS download only is optimal. The download cost is then
	\begin{equation}
		\label{eq:CdBS}
		\CdBS = M\omega\rhoBS.
	\end{equation}
\end{remark}

We also have the following result about the behavior of $\Cd$ in \eqref{eq:Cd}.

\begin{corollary}
\label{cor:dCd}
	For $\mu>0$, $\Cd$ is monotonically increasing with $\Delta$ if $\rho>\frac{h\alpha}{F}$, monotonically decreasing with $\Delta$ if $\rho<\frac{h\alpha}{F}$, and constant otherwise.
\end{corollary}

\begin{IEEEproof}
	The proof follows directly from differentiating $\Cd$ with respect to $\dr$ and is therefore omitted.
\end{IEEEproof}

\subsection{Overall Communication Cost}
\label{sec:TotalCost}

Combining Theorems \ref{thm:Cr} and \ref{thm:Cd}, one obtains the expression for the overall communication cost, 
\begin{align}
	\label{eq:C}
	\C = \Cr+\Cd.
\end{align}
Note that, in general, $\C$ is not monotone with $\dr$. We can derive the following result for $\dr= 0$ (instantaneous repair) and $\dr\rightarrow \infty$ (no repair).
\begin{corollary}
	\label{cor:Casym}
	\begin{equation}\label{eq:Cinst}
		\lim_{\dr\to0}\C = \frac{\rhoD}{F}(\gD m\mu+M\omega h\alpha).
	\end{equation}
	Moreover, for $\mu>0$,
	\begin{equation}
		\label{eq:Cno}
		\lim_{\dr\to\infty}\C = M\omega \rhoBS.
	\end{equation}
\end{corollary}
\begin{IEEEproof}
	See Appendix~\ref{prf:Casym}.
\end{IEEEproof}
For instantaneous repair ($\dr =0$), both repair and download are always performed from the storage nodes. Thus, the two terms in \eqref{eq:Cinst} correspond to the \ac{D2D} repair and \ac{D2D} download, and we recover the result in \cite{Paakkonen2013}. For $\dr \rightarrow \infty$, data is never repaired (hence, $\Cr=0$). For $\mu>0$, the number of storage nodes in the cell will become smaller than $h$ at some point, and \ac{D2D} download is no longer possible. Therefore, the overall communication cost in \eqref{eq:Cno}  is the \ac{BS} download cost in \eqref{eq:CdBS}.

\section{Hybrid Repair and Download}
\label{sec:hybrid}
In the system model in Section~\ref{sec:model} and the analysis in Section~\ref{sec:analysis} we assumed that if repair (resp. download) cannot be completed from storage nodes (because there are less than $r$ (resp. $h$) storage nodes available in the cell), \ac{BS} repair (resp. download) is performed. Alternatively, for both repair and download, a node might retrieve data from the available storage nodes using \ac{D2D} communication and retrieve the rest from the \ac{BS} to complete the repair or the download. 
We will refer to this setup as partial \ac{D2D} repair and partial \ac{D2D} download, and the scheme that implements it as the \emph{hybrid repair and download scheme}. In the following, we extend the analysis in Section~\ref{sec:analysis} to the hybrid scheme.

\subsection{Repair Cost}

Assume that, at the time of repair, $i<r$ storage nodes are available, i.e., repair cannot be accomplished exclusively from the storage nodes. However, $i\beta$ bits could be retrieved from the $i$ available storage nodes and the remaining $\gD-i\beta=(r-i)\beta$ bits to complete the repair from the \ac{BS}. The corresponding communication cost is $(\rhoBS(r-i)+\rhoD i)\beta$. For the conventional scheme, \ac{D2D} repair is not possible for $i<r$, and the repair cost corresponds to that of \ac{BS} repair, i.e., $\rhoBS\gBS$. This implies that, if $i<r$, partial repair leads to a reduced repair cost if $(\rhoBS(r-i)+\rhoD i)\beta<\rhoBS\gBS$ or, equivalently, $i>\frac{\rhoBS}{\rhoBS-\rhoD}\left(r-\frac{\gBS}{\beta}\right)\triangleq \cc$. 
For $i<r$, the hybrid scheme performs partial \ac{D2D} repair if $i>\cc$ and \ac{BS} repair otherwise. The repair cost is given in the following theorem.
\begin{theorem}
\label{lem:hbrCr}
Consider the DS network in Section~\ref{sec:model} using the hybrid scheme. The repair cost is given by
\begin{align*}
	\Crh &= \frac{1}{F\dr}\Bigg(\rhoBS\gBS\sum_{i=0}^{a}(m-i)b_i(m,p)\\
		&+\sum_{i=a+1}^{r-1}(m-i)(\rhoBS(r-i)+i\rhoD)\beta b_i(m,p)\\
		&+\rhoD\gD\sum_{i=r}^m(m-i)b_i(m,p)\Bigg),
\end{align*}
where $a = \min\left\{\left\lfloor \frac{\rhoBS}{\rhoBS-\rhoD}\left(r-\frac{\gBS}{\beta}\right)\right\rfloor,r-1\right\}$, $\left(r-\frac{\gBS}{\beta}\right)\ge 0$ for all codes in Section~\ref{sec:codes}, and $p=e^{-\mu\dr}$.
\end{theorem}
\begin{IEEEproof}
	It follows the same lines as the proof of Theorem~\ref{thm:Cr}.
\end{IEEEproof}

\subsection{Download Cost}


Similar to the repair case, if $i<h$ storage nodes are available at the time of a file request, the file cannot be downloaded solely from the storage nodes. However, $i\alpha$ bits could be downloaded from the $i$ available storage nodes and the remaining $(h-i)\alpha$ bits from the \ac{BS}, with communication cost $(\rhoBS(h-i)+\rhoD i)\alpha$. For the conventional scheme, the download cost corresponds to that of \ac{BS} download, i.e., $\rhoBS F$. Hence, the hybrid scheme leads to a lower download cost if $(\rhoBS(h-i)+\rhoD i)\alpha<\rhoBS F$, or equivalently, $i>\frac{\rhoBS}{\rhoBS-\rhoD}\left(h-\frac{F}{\alpha}\right)\triangleq \dd$.
For $i<h$, the hybrid scheme performs partial \ac{D2D} download if $i>\dd$ and \ac{BS} download otherwise. The download cost is given in the following theorem.

\begin{theorem}
\label{lem:Cdhbr}
	Consider the DS network in Section~\ref{sec:model} using the hybrid scheme. Let $\mu_i=i\mu$ and $p_i=e^{-\mu_i\dr}$, for $i=1,\ldots,m$. The download cost is given by
	\begin{align}
		\Cdh &= \frac{M\omega}{F}\Bigg(\rhoBS F\Bigg(1-\frac{1}{\dr}\sum_{i=1}^m\frac{1-p_{i}}{\mu_{i}}\prod_{\substack{j=1\\j\neq i}}^m\frac{j}{j-i}\Bigg) \nonumber\\
			&+\rhoBS F\sum_{i=1}^{a}\cc_i + \sum_{i=a+1}^{h-1}(\rhoBS(h-i)+i\rhoD)\alpha \cc_i \nonumber\\
			\label{eq:Cdhbr}
			&+\rhoD h\alpha\frac{1}{\dr}\sum_{i=h}^{m}\frac{1-p_i}{\mu_i}\prod_{\substack{j=h\\j\neq i}}^{m}\frac{j}{j-i}\Bigg),
	\end{align}
	where $a = \min\left\{\left\lfloor \frac{\rhoBS}{\rhoBS-\rhoD}\left(h-\frac{F}{\alpha}\right)\right\rfloor,h-1\right\}$, $\left(h-\frac{F}{\alpha}\right)\ge 0$, and
	\begin{align*}
		\cc_i 
		&= \frac{1}{\dr} \sum_{i'=i}^m\frac{1-p_{i'}}{\mu_{i'}}\prod_{\substack{j=i\\j\neq i'}}^m\frac{j}{j-i'}\\
			&-\frac{1}{\dr} \sum_{i'=i+1}^m \frac{1-p_{i'}}{\mu_{i'}}\prod_{\substack{j=i+1\\j\neq i'}}^m \frac{j}{j-i'}.
	\end{align*}

\end{theorem}
\begin{IEEEproof}
	See Appendix~\ref{prf:hbrdCd}.
\end{IEEEproof}

\section{Repair and Download Cost with\\ an Incoming Process}
\label{sec:Incoming}

The analysis in the preceding sections does not consider the possibility that nodes arriving to the cell may bring content. In a real scenario with neighboring cells, however, this may be the case. We will refer to the arrival of nodes with content as the \emph{incoming process}. Considering an incoming process significantly complicates the analysis. This is due to the fact that arriving nodes may bring content that is not \emph{directly useful}, in the sense that they may bring code symbols which are already available in another storage node. At a given time, it is likely that some symbols will be stored by more than one storage node, while other symbols will not be present in the storage network (due to node departures). As a result, the analysis needs to consider \emph{storage node classes}, where a node class defines the set of storage nodes storing given code symbols. 
In general, for an $[m,h,r]$ erasure correcting code, there are $m$ storage node classes, since all code symbols are different. The case of simply replicating the data (using a repetition code) is a bit different. Despite the fact that all code symbols are equal, for the analysis of $m$-replication we still need to consider $m$ storage node classes, i.e., we treat each of the code symbols of the $m$-replication as they were different.

In this section, we extend the analysis in Sections~\ref{sec:analysis} and \ref{sec:hybrid} to the scenario with an incoming process. In particular, we show that Theorem~\ref{thm:Cr} and Theorem~\ref{thm:Cd} can also be used to analyze the repair and download costs for this scenario by using different input parameters. More precisely, we consider the scenario where storage nodes of a given class arrive to the cell according to a Poisson process with expected arrival rate $\lambda_\text{c}\le \mu$.  
An incoming storage node brings a single code symbol of a given class. Furthermore, nodes not storing content arrive according to a Poisson process with expected arrival rate $M\lambda-m\lambda_\text{c}$. The departure rate for all nodes is $\mu=\lambda$, i.e., as before, the average number of nodes in the cell is $M$. We assume the practical scenario where the \ac{BS} maintains a list of the nodes storing content, which is communicated periodically to all nodes in the cell every $\Phi$ \acp{t.u.}. For simplicity, we assume that $\Phi=\Delta$.

\subsection{Repair Cost}

Denote by $c_{i}(t)$ the number of class-$i$ storage nodes in the cell at time $t$. Also, denote by $q_{i,0}(t)$ the probability that class $i$ is empty at time $t$, i.e., $q_{i,0}(t)=\Pr(c_{i}(t)= 0)$. Since all storage node classes have the same arrival and departure rate, we can drop subindex $i$ and write $q_{i,0}(t)=q_0(t)~\forall i$. Also, let $\q=(q_0,q_1,\ldots)$ be the stationary distribution, where $q_j$ is the probability that class $i$ has $j$ storage nodes. Equation \eqref{eq:CrT} in Theorem~\ref{thm:Cr} can then be used for the scenario with an incoming process by setting $p\leftarrow (1-q_0)$.

The difficulty here lies in computing $\q$. Without repairs, the evolution of $c_{i}(t)$ is given by a Poisson birth-death process, which can be modeled by an $\mathsf{M}/\mathsf{M}/\infty$ Markov chain model. In this case, the stationary distribution $\q=(q_0,q_1,\ldots)$ exists and can be computed. However, the repairs performed every $\Delta$ t.u. interfere with the stationarity of the process. Indeed, in the presence of repairs, the evolution of $c_{i}(t)$ does no longer correspond to a Poisson birth-death process. In this case, the analysis appears to be formidable. 

Here, we propose the following two-step procedure to compute $\q$. Consider a single repair interval of duration $\Delta$, where $c_i(\Delta)$ is the number of storage nodes in class $i$ at time $\Delta$. Within a repair interval $t\in[0,\Delta)$, $c_i(t)$ is described by a  Poisson birth-death process\footnote{This is contrast to the case with no incoming process, where the evolution of $c_i(t)$ for $t\in[0,\Delta)$ is described by a Poisson death process.}. Since  storage node classes are independent of each other and have the same arrival and departure rates, we can focus on a single class. Hence, we will drop the subindex $i$ in $c_i(t)$ and simply write $c(t)$. 

Let $P_{ij}(t)=\Pr(c(t)=j|c(0)=i)$ denote the transition probability function of the continuous-time $\mathsf{M}/\mathsf{M}/\infty$ Markov chain representing the Poisson birth-death process. $P_{ij}(t)$ can be computed by deriving a set of differential equations, called Kolmogorov's forward equations, whose solution can be computed as follows \cite{Ste09}. Let $\P(t)$ be the $S\times S$ matrix with $(i,j)$th entry $P_{ij}(t)$, where $S-1$ is the maximum number of storage nodes of one class. Also, let $r_{ij}$ be the transition rates of the continuous-time Markov chain. Then $\P(t)$ can be computed as \cite{Ste09}
\begin{align}
	\label{eq:KolmogorovSolution}
	\P(t)=e^{t\G}\triangleq\sum_{\ell=0}^\infty\frac{(t\G)^{\ell}}{\ell!},
\end{align}
where $\G$ is the generator of the Markov chain, with 
entries $g_{ij}$, $i=0,\ldots,S-1$ and $j=0,\ldots,S-1$, given by
\begin{align*}
g_{ij}&=r_{ij}~~\text{for}~i\neq j,\\
g_{ii}&=-\sum_{j=0}^{S-1}r_{ij},
\end{align*}
with
\begin{align}
r_{ij}=\left\{\begin{array}{cl}\lambda_\text{c} &  \text{if}~j=i+1 \\ i\mu & \text{if}~j=i-1\\\ 0 & \text{otherwise}   \end{array}\right. .
\end{align}

The infinite power series in \eqref{eq:KolmogorovSolution} converges for any square matrix $\G$, and can be efficiently computed using, e.g., the algorithm described in \cite{AlM09}.


Note that in our scenario, $S$ is not finite. However, if $\lambda_\text{c}\le \mu$ the probability of having $c(t)=j$ storage nodes of a given class at time $t$, $\Pr(c(t)=j)$, sharply decreases with $j$. Therefore, we can limit $S$ to a sufficiently large value, and by solving \eqref{eq:KolmogorovSolution} get a very good approximation of $\P(\Delta)$.

Given $\P(\Delta)$, we can estimate the stationary distribution $\q$ recursively. For a given distribution at time $t=0$, $\q(0)$, we can compute $\q(i\Delta)$ as
\begin{align}
	\label{eq:rec}
	q_j(i\Delta)=\sum_{\ell=0}^{S-1} P_{\ell j}(\Delta) \tilde{q}_{\ell}((i-1)\Delta),~~j=0,\ldots,S-1,
\end{align}
where $\tilde{q}_0(i\Delta)=0$ and $\tilde{q}_1(i\Delta)=q_0(i\Delta)+q_1(i\Delta)$, due to the repair, and $\tilde{q}_{\ell}(i\Delta)=q_{\ell}(i\Delta)$ for $\ell=2,\ldots,S-1$.

Equivalently, this recursion can be written in compact form as
\begin{align}
\label{eq:rec1}
\tilde{\q}&=\lim_{N\rightarrow\infty}\q(0)(\P(\Delta)\X)^{N},\\
\label{eq:rec2}
\q&=\tilde{\q}\P(\Delta),
\end{align}
where $\X$ is an $S\times S$ matrix with entries $x_{00}=0$, $x_{ii}=1$ for $i>0$, and $x_{01}=1$. Note that $\q$ and $\tilde{\q}$ are the stationary distributions before and after repair, respectively.



\begin{theorem}
\label{thm:CrIncoming}
	Consider the DS network in Section~\ref{sec:model} with departure rate $\mu$, arrival rate of storage nodes of a given class $\lambda_\textnormal{c}$, arrival rate of nodes not storing content $M\lambda-m\lambda_\textnormal{c}$, communication costs $\rhoBS$ and $\rhoD$, BS repair bandwidth $\gBS$, file size $F$, and repair interval $\Delta$. Furthermore, consider the use of an $[m,h,r]$ erasure correcting code with D2D repair bandwidth $\gD$. The repair cost is given by \eqref{eq:CrT} with $p\leftarrow(1-q_0)$, and $q_0$ is given by the first element of $\q$ in \eqref{eq:rec2}. 
\end{theorem}
\begin{IEEEproof}
	The proof follows from the discussion above.
\end{IEEEproof}

\begin{theorem}
\label{lem:hbrCrIncoming}
Consider the DS network in Section~\ref{sec:model} with departure rate $\mu$, arrival rate of storage nodes of a given class $\lambda_\textnormal{c}$,  and arrival rate of nodes not storing content $M\lambda-m\lambda_\textnormal{c}$, using the hybrid scheme. The repair cost is given by the expression in Theorem~\ref{lem:hbrCr} with $p\leftarrow(1-q_0)$, and $q_0$ is given by the first element of $\q$ in \eqref{eq:rec2}. 
\end{theorem}
\begin{IEEEproof}
	The proof follows from the discussion above.
\end{IEEEproof}

\begin{remark}
It is important to remark that the analysis for the scenario with an incoming process does not consider the departure of individual storage nodes, but rather the departure of whole classes, i.e., all nodes of a given class. Thus, $r$ and $m$ in \eqref{eq:CrT} should not be interpreted as $r$ storage nodes and $m$ storage nodes, respectively, but as $r$ and $m$ storage node classes.
\end{remark}

\begin{remark}
Note that in the analysis above we have made the assumption that the stationary distribution $\q$ exists. While we do not have a formal proof for this, our numerical results suggest that it does exist. In fact, the recursion \eqref{eq:rec1} and \eqref{eq:rec2} converges to the same $\q$ independently of $\q(0)$.  
\end{remark}

\subsection{Download Cost}
\label{sec:DownloadCostIncoming}

Assume that after repair there are $\ell$ storage nodes of a given class, say class $i$. With some abuse of notation, let $c_{i}(t,\ell)$ be the number of storage nodes of class $i$ at time $t$, where parameter $\ell$ indicates that $c_{i}(t=0)=\ell$. The evolution of $c_{i}(t,\ell)$ for $t\in[0,\Delta)$ is given by a Poisson death process.  Denote by $U_{\ell}$ the time instant at which the last of the $\ell$ storage nodes in class-$i$ leaves the cell. $U_{\ell}$ is hypoexponentially distributed with pdf given by \eqref{eq:Shpdf}, with $h=1$ and $m={\ell}$. The expected value of $U_{\ell}$ is \cite[Sec.~1.3.1]{Bolch2006}
\begin{align}
	\mathbb{E}[U_{\ell}]=\sum_{j=1}^{\ell}\frac{1}{j\mu}.
\end{align}
Note that $U_1$ is exponentially distributed.

Let $U$ be the time instant at which the last of the storage nodes in class $i$ leaves the cell or, in other words, the time instant at which the whole class $i$ leaves the cell. The pdf of $U$ is a weighted sum of the pdfs $U_{\ell}$, weighed by $\tilde{q}_{\ell}$, i.e., it is a weighted sum of hypoexponential distributions. The expected value of $U$ is  
\begin{align}
	\mathbb{E}[U]=\sum_{\ell=0}^{\infty}\tilde{q}_{\ell}\sum_{j=1}^{\ell}\frac{1}{j\mu}.
\end{align}

Let $b(t)$ be the number of nonempty storage node classes in the cell at time $t$. Computing $\Cd$ exactly requires to compute the distribution of the time instant at which $b(t)$ changes from $h$ to $h-1$, denoted by $S_h$, similar to the case with no incoming process (see Appendix~\ref{prf:Cd}). Unfortunately, due to the fact that the pdf of $U$ is a weighted sum of hypoexponential distributions, computing the pdf of $U$ seems unfeasible. Here, we propose to approximate the pdf of $U$ by an exponential pdf. Indeed, it appears that $\tilde{q}_1$ is in general the largest element in $\tilde{\q}$, therefore the distribution of $U$ has a large exponential component. Assuming that $U$ is well approximated by an exponential distribution with mean $\tilde{\mu}^{-1}$, the download cost for the scenario with an incoming process can then be computed using \eqref{eq:Cd} in Theorem~\ref{thm:Cd} by setting $\mu\leftarrow\tilde{\mu}$, where now a storage node departure should be interpreted as a storage node class departure. We have observed that by approximating the pdf of $U$ by an exponential distribution with mean $\tilde{\mu}^{-1}=\mathbb{E}[U]$, the analytical results match very well with the simulations for the whole range of interesting values of $\mu$ and $\lambda_\text{c}$, as shown in the results section. The download cost for the hybrid scheme is found by using \eqref{eq:Cdhbr} in Theorem~\ref{lem:Cdhbr} with $\mu\leftarrow \tilde{\mu}$.


\section{Erasure Correcting Codes in Distributed Storage}
\label{sec:codes}
From Sections~\ref{sec:analysis}--\ref{sec:Incoming}, it can be seen that the overall communication cost $\C$ depends on the network parameters $\mu$ ($\lambda$), $\lambda_\text{c}$, $M$, and $\omega$, and on the parameters $m$, $h$, $r$, $\alpha$, and $\beta$ (and subsequently on $\gD=r\beta$ and $\gBS=\alpha$), which are determined by the erasure correcting code used for \ac{DS}.
An erasure correcting code for \ac{DS} is typically described in terms of the number of nodes used for storage, the download locality and the repair locality, and is defined using the notation $[m,h,r]$. In this section, we briefly describe \ac{MDS} codes \cite{Ryan2009}, regenerating codes \cite{Dimakis2010} and \acp{LRC} \cite{Papailiopoulos2012} in the context of \ac{DS}. We also connect the code parameters $[m,h,r]$ with the code parameters $(n,k)$.  In Section~\ref{sec:results}, we then evaluate the overall communication cost of \ac{DS} using these three code families. 

We remark that the analysis in the previous sections applies directly to MDS and regenerating codes. However, due to the specificities of LRCs, Theorem~\ref{thm:Cr} needs to be slightly modified, as shown in Section~\ref{sec:LRC} below.

\subsection{Maximum Distance Separable Codes}\label{sec:mds}
Assume the use of an $(n,k)$ \ac{MDS} code for \ac{DS}. In this case, each storage node stores one coded symbol, hence $m=n$ and $\aMDS=\frac{F}{k}$. Due to the \ac{MDS} property, \ac{D2D} repair and \ac{D2D} download require to contact $r=h=k$ storage nodes. Therefore, an $(n,k)$ MDS code in a \ac{DS} context is described with the triple $[n,k,k]$. Moreover, $\bMDS=\aMDS=\frac{F}{k}$, i.e., $\gD=F$. The fact that an amount of information equal to the size of the entire file has to be retrieved to repair a single storage node is a known drawback of \ac{MDS} codes \cite{Dimakis2010}. The simplest \ac{MDS} code is the $n$-replication scheme. In this case, each storage node stores the entire file, i.e., $\arep=F$ and $r=h=k=1$.

\subsection{Regenerating Codes}
\label{sec:regenerating}
A lower repair bandwidth $\gD$ (as compared to \ac{MDS} codes) can be achieved by using regenerating codes \cite{Dimakis2010}, at the expense of increasing $r$ \cite{Dimakis2010}. Two main classes of regenerating codes are covered here, \ac{MSR} codes and \ac{MBR} codes. \ac{MSR} codes yield the minimum storage per node, i.e., $\aMSR$ is minimum, while \ac{MBR} codes achieve minimum \ac{D2D} repair bandwidth. Regenerating codes have two repair models, \emph{functional repair} and \emph{exact repair} \cite{Dimakis2011}. In exact repair, the lost data is regenerated exactly \cite{Dimakis2011}. In functional repair, the lost data is regenerated such that the initial state of reliability in the \ac{DS} system is restored \cite{Dimakis2011}, but the regenerated data does not need to be a replica of the lost data \cite{Dimakis2011}. Here, we consider only exact repair, since it is of more practical interest \cite{Rashmi2011}.

An exact-repair $[m,h,r]$ \ac{MSR} code in a \ac{DS} system has $k = h(r-h+1)$ and $n=m(r-h+1)$, with $r=2(h-1),\ldots,m-1$ \cite{Rashmi2011}.\footnote{The design of linear, exact-repair \ac{MSR} codes with $r<2(h-1)$ has been proven impossible \cite{Shah2012}.} Hence, using \eqref{eq:alpha},
\begin{equation*}
	\aMSR = \frac{1}{m}\cdot\frac{F}{R} = \frac{F}{m}\cdot\frac{m(r-h+1)}{h(r-h+1)} = \frac{F}{h}.
\end{equation*}
Furthermore \cite{Rashmi2011},
\begin{equation*}
	\bMSR=\frac{F}{k}=\frac{F}{h} \cdot \frac{1}{r-h+1}\le\aMSR,
\end{equation*}
with equality only when $r=h$, which is only possible for $h=1$ and $h=2$ due to the restriction on the values for the repair locality. The repair bandwidth,
\begin{equation*}
	\gD=r\bMSR=\frac{F}{h} \cdot \frac{r}{r-h+1}\le F,
\end{equation*}
is minimized for $r=m-1$ \cite{Dimakis2010}. We remark that the storage per node $\alpha$ (and hence the average download cost) for an $(m,h)\equiv [m,h,h]$ \ac{MDS} code and an $[m,h,r]$ \ac{MSR} code are equal.

An \ac{MBR} code further reduces the repair bandwidth at the expense of increasing the storage per node. An exact-repair $[m,h,r]$ \ac{MBR} code has $k = hr-\binom{h}{2}$ and $n=mr$ for $r=h,\ldots,m-1$ \cite{Rashmi2011}. Using \eqref{eq:alpha}, we have
\begin{equation*}
	\aMBR = \frac{1}{m}\cdot\frac{F}{R} = \frac{F}{m}\cdot\frac{2mr}{h(2r-h+1)} = \frac{F}{h}\cdot\frac{2r}{2r-h+1}.
\end{equation*}
Furthermore \cite{Rashmi2011},
\begin{equation*}
	\bMBR = \frac{F}{k} = \frac{F}{h} \cdot \frac{2}{2r-h+1} \le \aMBR.
\end{equation*}
Similar to the \ac{MSR} codes, the repair bandwidth of an \ac{MBR} code,
\begin{equation*}
	\gD=r\bMBR=\frac{F}{h}\frac{2r}{2r-h+1}\le F,
\end{equation*}
is minimized for $r=m-1$ \cite{Dimakis2010}.

Note that an $[m,1,r]$ regenerating code has exactly the same overall communication cost as an $m$-replication scheme.

\subsection{Locally Repairable Codes}
\label{sec:LRC}
A lower repair locality $r$ (as compared to \ac{MDS} codes) is achieved by using \acp{LRC} \cite{Papailiopoulos2012}. An $[m,h,r]$ \ac{LRC} has $k = rh$ and $n = m(r+1)$, where $r<h$ and $(r+1)~|~m$. Each node stores
\begin{equation*}
	\aLRC = \frac{1}{m}\cdot\frac{F}{R} = \frac{F}{m} \cdot \frac{m(r+1)}{rh} = \frac{F}{h} \cdot \frac{r+1}{r}
\end{equation*}
bits. The storage nodes are arranged in $G\triangleq\frac{m}{r+1}$ disjoint repair groups with $r+1$ nodes in each group. Any single storage node can be repaired \emph{locally} by retrieving $\gD=r\bLRC$ bits from $r$ nodes in the repair group \cite{Papailiopoulos2012}. A storage node involved in the repair process transmits all its stored data, i.e., $\bLRC=\aLRC$, hence
\begin{equation*}
	\gD=r\bLRC=\frac{F}{h}(r+1)\le F.
\end{equation*}

If \textit{local} \ac{D2D} repair is not possible, repair can be carried out \emph{globally} by retrieving $h\aLRC$ bits from any subset of $h$ storage nodes. Since it is necessary to distinguish between local and global repairs (as opposed to MDS and regenerating codes), the expression of the repair cost $\Cr$ in Theorem~\ref{thm:Cr} does not apply to LRCs and needs to be modified. We denote by $\mrlD$ and $\mrgD$ the average number of nodes repaired from the storage nodes locally and globally, respectively, in one repair interval. We will also need the following definitions. Let $\X\triangleq(X_0,X_1,\ldots,X_{r+1})$ be the random vector whose component $X_i$ is the random variable giving the number of repair groups with $i$ storage node departures in a repair interval $\dr$. Note that $X_i$ takes values in $\{0,1,\ldots,G\}$ and $\sum_iX_i=G$. The probability of $i$ storage node departures in a repair group is $y_i\triangleq\binom{r+1}{i}p^{r+1-i}(1-p)^i$, where $p=e^{-\mu\dr}$ is the probability that a storage node has not left the network during a time $\dr$. Let $\x\triangleq(x_0,x_1,\ldots,x_{r+1})$ be a realization of $\X$ and let $\y\triangleq(y_0,y_1,\ldots,y_{r+1})$. Then, 
\begin{align}
	\Pr(\X=\x) & = \sum_{\x:|\x|=G}\binom{G}{\x}\y^{\x},
	\label{eq:multpmf}
\end{align}
where $|\x|\triangleq\sum_i x_i$, $\binom{G}{\x}\triangleq\frac{G!}{x_0!x_1!\cdots x_{r+1}!}$ is the multinomial coefficient, and $\y^{\x} \triangleq \prod_iy_i^{x_i}$.

The repair cost for LRCs is given in the following theorem.
\begin{theorem}
\label{thm:LRCrep}
Consider the DS network in Section~\ref{sec:model} with departure rate $\mu$, communication costs $\rhoBS$ and $\rhoD$, BS repair bandwidth $\gBS$, file size $F$, and repair interval $\Delta$. Furthermore, consider the use of an $[m,h,r]$ LRC with $G$ disjoint repair groups and D2D repair bandwidth $\gD$. The repair cost is given by
\begin{equation}
\label{eq:CrLRC}
	\Cr = \frac{1}{F\dr}\left( \rhoBS\gBS \mrBS+\rhoD\left(\gD\mrlD+h\aLRC\mrgD\right) \right),
\end{equation}
where
\begin{align*}
	\mrlD & = mp^r(1-p),\\
	\mrgD & = \sum_{\x:|\x|=G}\binom{G}{\x}\y^{\x} \cdot \sum_{i=2}^{r+1}ix_i \cdot \mathds{1}\left\{\sum_{i=1}^{r+1}ix_i\le m-h\right\},\\
	\mrBS & = m(1-p) - \mrlD -\mrgD,
\end{align*}
$p=e^{-\mu\dr}$ and $\mathds{1}\{\cdot\}$ is an indicator function.
\end{theorem}
\begin{IEEEproof}
	See Appendix~\ref{prf:LRCrep}.
\end{IEEEproof}
\noindent It is easy to verify that Corollary~\ref{cor:Casym} holds also for \acp{LRC}.

\subsection{Lowest Overall Communication Cost for Instantaneous Repair}

For instantaneous repair, the minimum overall communication cost is given in the following lemma.
\begin{lemma}
	\label{lem:optcode}
	For $\dr=0$ (instantaneous repair), the lowest possible overall communication cost for any $[m,h,r]$ linear code with $m=n$, regenerating codes and LRCs is
	\begin{equation*}
		\C_{\min}({\Delta=0})\triangleq \min_{m,h,r} \lim_{\Delta\rightarrow 0}\C=\rhoD(2\mu+M\omega),
	\end{equation*}
where $\lim_{\dr\rightarrow 0}\C$ is given in \eqref{eq:Cinst} in Corollary~\ref{cor:Casym}. The minimum is achieved by $2$-replication.
\end{lemma}
\begin{IEEEproof}
	See Appendix~\ref{prf:optcode}.
\end{IEEEproof}
This is in agreement with the result in \cite{Paakkonen2013}, where $2$-replication was shown to be optimal.


\section{Numerical results}
\label{sec:results}
\begin{figure}[!t]
\begin{center}
	\includegraphics[width=1\columnwidth]{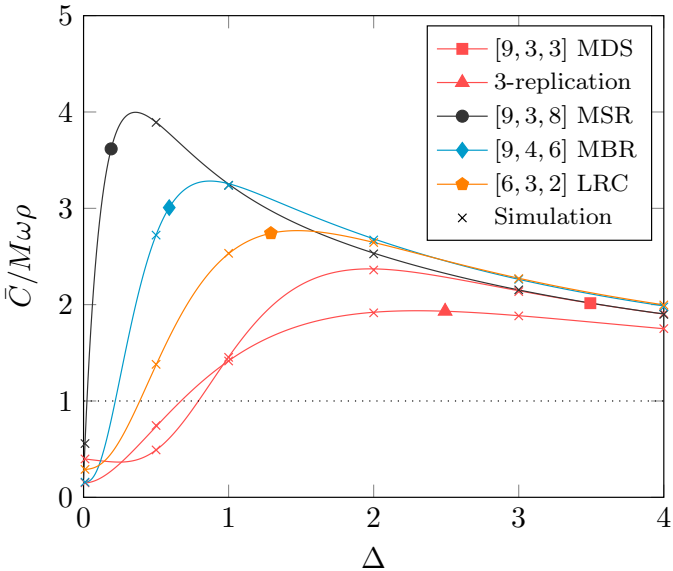}
	\vspace{-4ex}
	\caption{Normalized overall communication cost $\C/M\omega\rho$ versus the repair interval $\dr$ for a selection of \ac{MDS} codes, regenerating codes and \acp{LRC} with $R=1/3$, compared to the normalized \ac{BS} download cost (dotted line).}
	\label{fig:cost}
	\vspace{-3ex}
\end{center}
\end{figure}

In this section, we evaluate the overall communication cost $\C$ (computed using \eqref{eq:CrT} and \eqref{eq:Cd}) for the erasure correcting codes discussed in the previous section. For the results, we consider a network with $M=30$ nodes, where the number of storage nodes is $m\le10$. This gives a probability smaller than $7.2\cdot10^{-6}$ of having less than $m$ nodes in the cell (see \eqref{eq:empty}), which is considered negligible. Without loss of generality, we set the departure rate $\mu=1$ and $\rhoD=1$, i.e., $\rho=\rhoBS$. Figs.~\ref{fig:cost}--\ref{fig:minG3w1rho40} refer to a system with no incoming process, i.e., $\lambda_\text{c}=0$, while Figs.~\ref{fig:costIncomA} and \ref{fig:costIncomB} consider the presence of an incoming process, $\lambda_\text{c}\ge0$.


Fig.~\ref{fig:cost} shows $\C$ normalized to the cost of downloading from the \ac{BS}, $M\omega\rho$, i.e., $\C/M\omega\rho$, as a function of the normalized repair interval, $\mu\dr=\dr$, for a selection of \ac{MDS} codes, regenerating codes and \acp{LRC} with $R=1/3$. The ratio between the request rate and departure rate is $\omega/\mu = 0.02$, i.e., the average request rate in the cell is $M\omega=0.6$ requests per \ac{t.u.}, and $\rho=40$. The meaning of $\omega/\mu = 0.02$ is that each node places in average $0.02$ requests per node life time. Also, in the figure $\dr=1$ means that the repair interval is equal to one average node lifetime. Simulation results\footnote{When simulating the wireless \ac{DS} system, the repair process is not executed if the number of nodes in the cell is less than $m$ at the particular repair instant.} are also included in the figure (markers). Note that since we normalize $\C$ to the \ac{BS} download cost, values below ordinate $1$ correspond to the case where \ac{DS} is beneficial. For relatively high repair frequencies, all codes yield lower $\C$ than \ac{BS} download. However, $\C/M\omega\rho$ exceeds $1$, i.e., \ac{BS} download is less costly than the \ac{DS} communication cost, for values of the repair interval larger than a threshold, which we define as
\begin{equation}
	\label{eq:drmax}
	\drmax\triangleq\sup\left\{\dr:\C < M\omega\rho \right\}.
\end{equation}
For $\dr>\drmax$, retrieving the file from the \ac{BS} is always less costly, therefore storing data in the nodes is useless. $\drmax$ depends on the network parameters $M$, $\omega$, $\mu$ and $\rho$ as well as the code parameters $m$, $h$ and $r$. 

We see from Fig.~\ref{fig:cost} that the value of $\dr$ that minimizes $\C$, denoted by $\dropt$, depends on the code used for storage. In particular, $\dropt=0$ for the $[9,3,8]$ \ac{MSR} code, i.e., instantaneous repair is optimal. Performing an exhaustive search for $m\le10$, it is readily verified that the same is true for any of the codes in Section~\ref{sec:codes} with $r=m-1$. It is reasonable to assume that this will be the case also for $m>10$. On the other hand, $\dropt>0$ for the $[9,3,3]$ \ac{MDS} code. $\dropt$ depends on the network and code parameters. In particular, the tolerance to storage node departures in a repair interval affects $\dropt$. 
In Section~\ref{sec:varynetwork}, we investigate how the network parameters affect $\C$ and $\drmax$. In Section~\ref{sec:varycode}, we explore how the code parameters affect $\C$.


\subsection{Effect of Varying Network Parameters}
\label{sec:varynetwork}

\begin{figure}[!t]
\begin{center}
	\includegraphics[width=1\columnwidth]{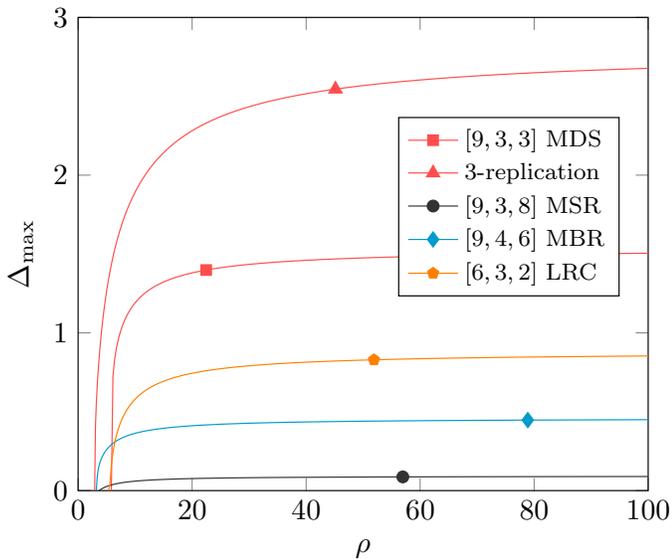}
	\vspace{-4ex}
	\caption{The maximum repair interval $\drmax$ versus the transmission cost ratio $\rho$.}
	\label{fig:drmaxrho}
	\vspace{-3ex}
	\end{center}
\end{figure}

Fig.~\ref{fig:drmaxrho} shows how $\drmax$ increases with $\rho$ for the same codes as in Fig.~\ref{fig:cost} and $\omega/\mu=0.05$. For $\rho<5$, approximately, $\drmax=-\infty$ for all considered codes, i.e., it is never beneficial to use the devices for storage and the file should always be downloaded from the \ac{BS}. It is worth noticing that, for moderate-to-large $\rho$, the $[9,3,8]$ \ac{MSR} code requires in the order of 10 repairs per average node lifetime while the $[9,3,3]$ \ac{MDS} code requires only around 0.66 repairs per node lifetime for \ac{DS} to be beneficial over \ac{BS} download. The main difference between the $[9,3,3]$ \ac{MDS} code and the $[9,3,8]$ \ac{MSR} code is the number of storage node departures in a repair interval that the code can tolerate such that \ac{D2D} repair is still possible, i.e., $m-r$. The $[9,3,3]$ MDS code can handle the departure of up to $6$ storage nodes while the $[9,3,8]$ \ac{MSR} code can tolerate a single departure only. This explains the higher repair frequency required by the MSR code.
\begin{figure}[!t]
\begin{center}
	\includegraphics[width=1\columnwidth]{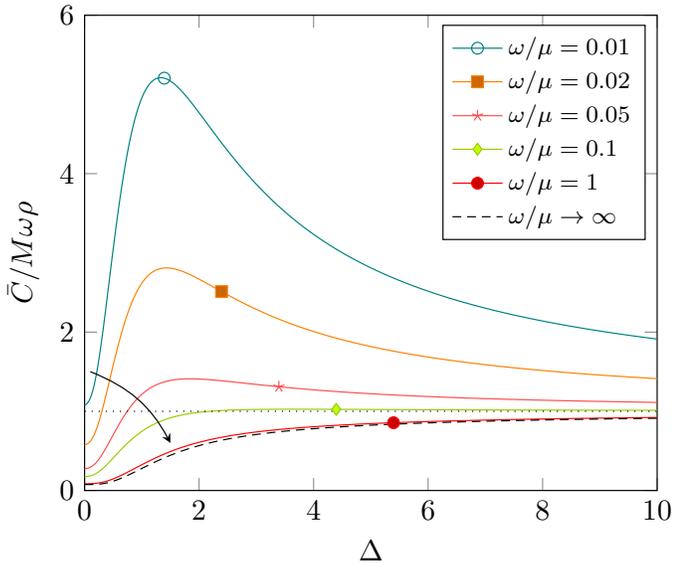}
	\vspace{-4ex}
	\caption{Normalized overall cost $\C/M\omega\rho$ versus the repair interval $\dr$ for the $[6,3,2]$ \ac{LRC} for different values of the ratio $\omega/\mu$, as compared with the normalized \ac{BS} download cost (straight dotted line). The arrow points in the direction of increasing $\omega/\mu$.}
	\label{fig:varw}
	\vspace{-3ex}
	\end{center}
\end{figure}

For the $[6,3,2]$ \ac{LRC} and $\rho=20$, Fig.~\ref{fig:varw} shows how $\C/M\omega\rho$ and $\drmax$ are affected by the ratio $\omega/\mu$.   We see that increasing $\omega/\mu$ reduces $\C/M\omega\rho$ for all $\dr$ and that $\drmax$ increases with $\omega/\mu$. The same behavior is observed using any of the codes in Section~\ref{sec:codes}, which can be verified by the following manipulations of the equations in Section~\ref{sec:analysis}. The case $\omega/\mu\to\infty$ corresponds to $\C/M\omega\rho\to\Cd/M\omega\rho$, which can be readily seen by taking the limit $\omega\to\infty$ in \eqref{eq:C}, using \eqref{eq:CrT} and \eqref{eq:Cd}, for fixed and finite $\mu$.  This shows that the overall communication cost is essentially the download cost for a sufficiently high $\omega/\mu$. Since $\Cd/M\omega\rho$ is monotonically increasing in $\dr$ (Corollary~\ref{cor:dCd}) and $\C/M\omega\rho\to1$ as $\dr\to\infty$ (Corollary~\ref{cor:Casym}), we also have that $\drmax\to\infty$ for $\omega/\mu\to\infty$. Hence, \ac{DS} always leads to a lower overall communication cost, as compared to the \ac{BS} download cost, for sufficiently large $\omega/\mu$.


\subsection{Results of Changing Code Parameters}
\label{sec:varycode}
\begin{figure}[!t]
\begin{center}
	\includegraphics[width=1\columnwidth]{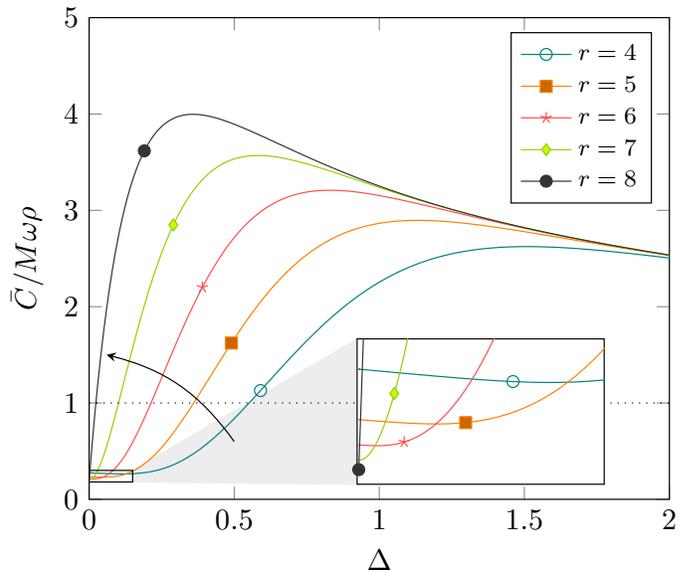}
	\vspace{-4ex}
	\caption{Normalized overall cost $\C/M\omega\rho$ versus the repair interval $\dr$ for the $[9,3,r]$ \ac{MSR} code compared with the normalized \ac{BS} download cost (dotted line). The arrow shows the direction of increasing $r$.}
	\label{fig:varr}
	\vspace{-2ex}
	\end{center}
\end{figure}

We investigate how the repair locality $r$ affects $\C$. Fig.~\ref{fig:varr} shows $\C/M\omega\rho$ versus $\dr$ for the $[9,3,r]$~\ac{MSR} code for $\rho=40$ and $\omega/\mu=0.02$. We observe that for $\dr=0$ the lowest $\C$ is achieved for $r=8$, i.e., the highest possible repair locality. This is due to the fact that for regenerating codes $\gD$ is minimized for $r=m-1$ (see \cite{Dimakis2010} and Section~\ref{sec:regenerating}). However, increasing $\dr$ requires decreasing $r$ to yield the lowest $\C$. This is due to the improved tolerance to storage node departures as $r$ decreases. The result is interesting, because it means that in wireless \ac{DS}, if repairs cannot be accomplished very frequently, repair locality is a more important parameter than repair bandwidth. On the other hand, if repairs can be performed very frequently, repair bandwidth becomes more important than repair locality, because tolerance to storage node departures is not critical. In general, there is a tradeoff between the repair bandwidth and the tolerance to storage node departures (directly related to the repair locality), which holds true for any of the codes in Section~\ref{sec:codes}. 
How to set the the parameter $r$ depends on how frequently we can repair the \ac{DS} system.

\subsection{Improved Communication Cost Using the Hybrid Scheme}
\begin{figure}[!t]
\begin{center}
	\includegraphics[width=1\columnwidth]{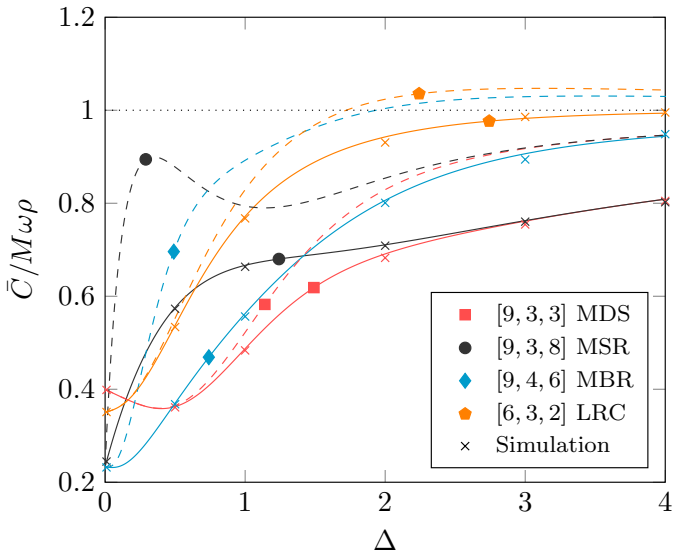}
	\vspace{-4ex}
	\caption{The normalized overall cost $\C/M\omega\rho$ versus the repair interval $\dr$ when using the conventional scheme (dashed curves) and hybrid scheme (solid curves).}
	\label{fig:hybrid}
	\vspace{-3ex}
	\end{center}
\end{figure}

We return to the hybrid repair and download scheme presented in Section~\ref{sec:hybrid} to investigate the gains in overall communication cost as compared to the cost when using the conventional scheme. We remark that the hybrid scheme does not improve $\C$ for all codes in Section~\ref{sec:codes}. In particular, for finite $\rho$, $\Cr$ is only reduced if $\beta<\alpha$ (Theorem~\ref{lem:hbrCr}) and $\Cd$ is only improved if $\alpha<F$ (Theorem~\ref{lem:Cdhbr}). Fig.~\ref{fig:hybrid} shows $\C/M\omega\rho$ versus $\dr$ for all codes in Fig.~\ref{fig:cost} that achieve lower $\C$ when using the hybrid scheme. We set $\omega/\mu=0.1$ and $\rho=10$ and include simulation results in the figure (markers). Dashed curves correspond to the conventional scheme, and solid curves to the hybrid scheme.
We see from the figure that regenerating codes achieve a large cost reduction, especially for small $\dr$, when using the hybrid scheme. This is since both $\Cr$ and $\Cd$ are reduced. A smaller cost reduction is observed for \ac{MDS} codes and \acp{LRC}.

\subsection{Codes Achieving Minimum Cost for Given $\dr$}
\label{sec:mincost}

The analytical expressions for the overall communication cost derived in Sections~\ref{sec:analysis} and \ref{sec:hybrid} can be used to find, for a given repair interval, the code achieving the lowest $\C$.
We have performed an exhaustive search for all \ac{MDS} codes (including replication), regenerating codes and \acp{LRC}, with $m\le10$, to find the code achieving the lowest $\C$ for each $\dr$. Like \cite{Leong2012}, we also introduce an overall storage budget constraint of $\Gamma$ files ($\Gamma F$ bits) across the nodes in the cell, i.e., $m\alpha\leq \Gamma F$. In particular, we set $\Gamma=3$, meaning that the code rate is $R\ge1/3$.

\begin{figure}[!t]
\begin{center}
	\includegraphics[width=1\columnwidth]{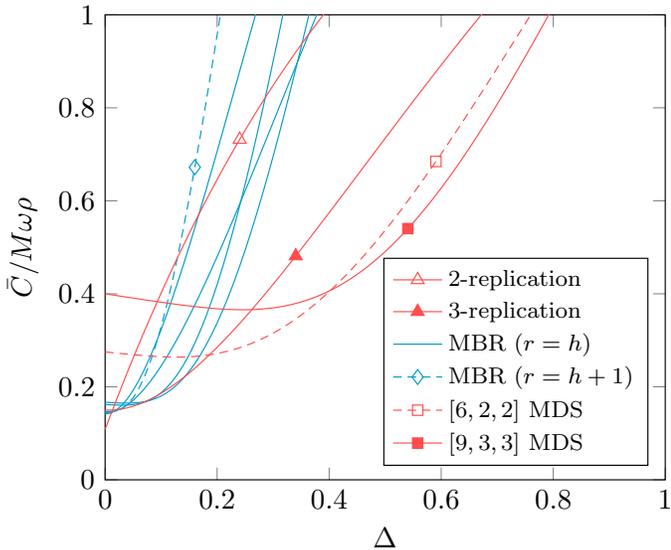}
	\vspace{-4ex}
	\caption{Codes achieving minimum $\C$ for some $\dr$ for $\omega/\mu=0.02$, $\rho=40$, and $\Gamma=3$.}
	\label{fig:minG3w002rho40}
	\vspace{-3ex}
	\end{center}
\end{figure}

Fig.~\ref{fig:minG3w002rho40} shows $\C/M\omega\rho$ for all codes that entail the lowest $\C$ for some value of $\dr$ for $\omega/\mu=0.02$ and $\rho=40$. For $\dr=0$ (instantaneous repair) $2$-replication is optimal (see Lemma~\ref{lem:optcode}). However, $2$-replication remains optimal only if repair can be accomplished at least around 80 times per average node lifetime. For slightly larger $\dr$, \ac{MBR} codes achieve the lowest cost. It is worth stressing that the \ac{MBR} codes achieving the lowest $\C$ for some $\dr$ are characterized by a low repair locality ($r=h$ and $r=h+1$), i.e., fault tolerance to storage node departures to allow \ac{D2D} repair is more important than the repair bandwidth. Somewhat surprisingly, \ac{MDS} codes offer the best performance for higher $\dr$, despite the large $\gD$. We remark that LRCs are not optimal for any $\dr$ due to the poor tolerance to storage node departures in local \ac{D2D} repair and a larger $\alpha$ than \ac{MDS} codes for a given global tolerance to storage node departures. $\drmax\approx0.8$ is the largest $\dr$ such that \ac{DS} is beneficial over \ac{BS} download, using any of the codes in Section~\ref{sec:codes}.

\begin{figure}[!t]
\begin{center}
	\includegraphics[width=1\columnwidth]{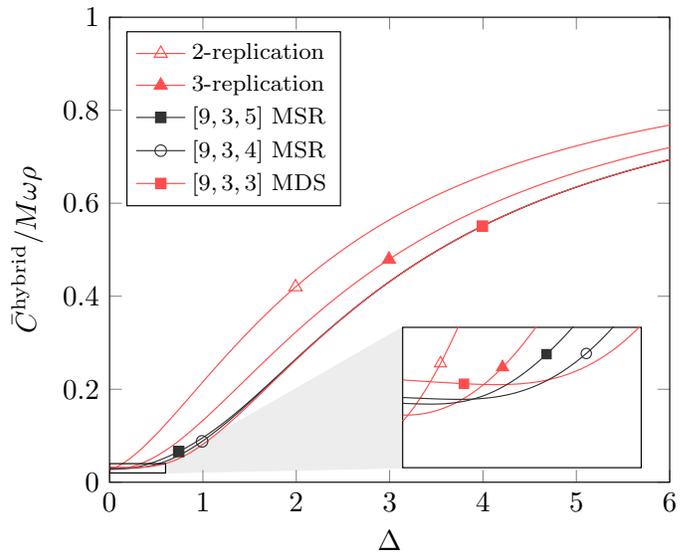}
	\vspace{-4ex}
	\caption{Codes achieving minimum $\Ch$ with the hybrid repair and download scheme for some $\dr$ when $\omega/\mu=1$, $\rho=40$, and $\Gamma=3$.}
	\label{fig:minG3w1rho40}
	\vspace{-3ex}
	\end{center}
\end{figure}

Fig.~\ref{fig:minG3w1rho40} shows the codes that achieve the lowest overall cost $\Ch=\Crh+\Cdh$ for some values of $\dr$ for the hybrid scheme with $\omega/\mu=1$ and $\rho=40$. Increasing $\omega/\mu$, $\Cd$ is the main contribution to $\C$ (see Section~\ref{sec:varynetwork}). Since $\alpha$ has significant impact on $\Cd$, we expect codes with small $\alpha$ to achieve the minimum cost. Indeed, we note that \ac{MDS} codes and \ac{MSR} codes, which have minimum $\alpha$, achieve the lowest $\C$ for a region of values of $\dr$. As expected, $2$-replication is optimal for instantaneous repair. 

\subsection{Scenario with an Incoming Process}
\label{sec:ResIncoming}

\begin{figure}[!t]
\begin{center}
	\includegraphics[width=1\columnwidth]{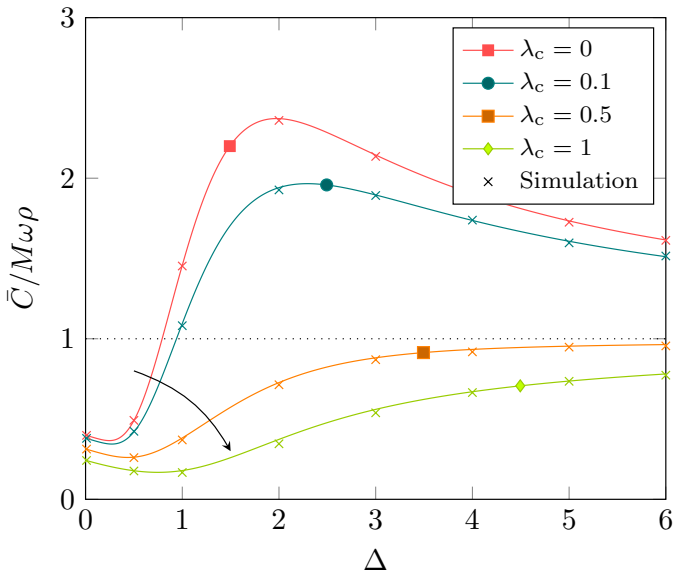}
	\vspace{-4ex}
	\caption{Normalized overall communication cost $\C/M\omega\rho$ versus the repair interval $\dr$ for the $[9,3,3]$ \ac{MDS} code for the scenario with an incoming process with different values of $\lambda_\text{c}$ for $\mu=1$, $\omega/\mu=0.02$, and $\rho=40$. The arrow shows the direction of increasing $\lambda_\text{c}$.}
	\label{fig:costIncomA}
	\vspace{-2ex}
\end{center}
\end{figure}

In Fig.~\ref{fig:costIncomA} we plot the analytical curves and simulation results for the $[9,3,3]$ MDS code for the scenario with an incoming process and several values of $\lambda_\text{c}$ when $\omega/\mu=0.02$ and $\rho=40$. The analytical curves for $\Cr$ (not shown here) match perfectly with the simulation results. However, as mentioned in Section~\ref{sec:DownloadCostIncoming}, to compute $\Cd$ we make the assumption that the pdf of the random variable representing the time instant at which the last of the storage nodes in a given class  leaves the cell is exponentially distributed. This translates into some slight discrepancies for $\Cd$, which obviously show also for $\C$. However, Fig.~\ref{fig:costIncomA} reveals a very good agreement between the analytical results and the simulation results, which justifies the assumption made. As expected, increasing $\lambda_\text{c}$ decreases the overall communication cost, since the average lifetime of a storage node class increases. Note that $\lambda_\text{c}=0$ corresponds to the case with no incoming process. For $\lambda_\text{c}=0.5$ and $\lambda_\text{c}=1$, where the latter corresponds to the realistic scenario where the arrival rate and departure rate of storage nodes is equal, wireless \ac{DS} is beneficial for any $\Delta$.

\begin{figure}[!t]
\begin{center}
	\includegraphics[width=1\columnwidth]{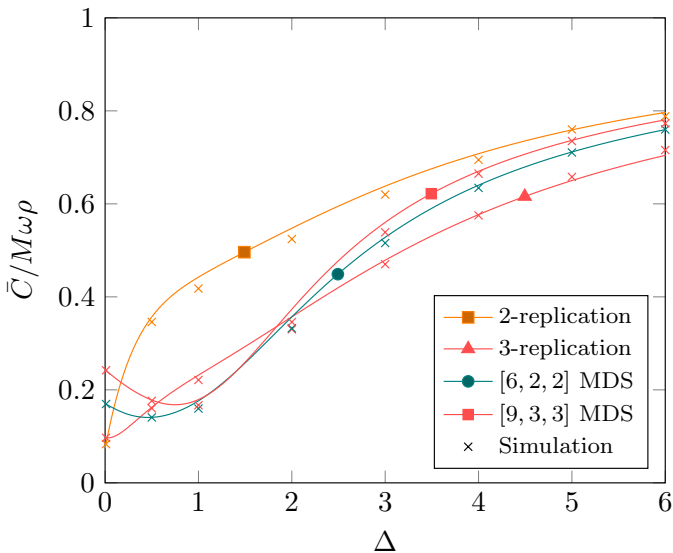}
	\vspace{-4ex}
	\caption{Codes achieving minimum $\C$ for some $\dr$ when $\lambda_\text{c}=\mu=1$, $\omega/\mu=0.02$, $\rho=40$, and $\Gamma=3$.}
	\label{fig:costIncomB}
	\vspace{-3ex}
\end{center}
\end{figure}

Fig.~\ref{fig:costIncomB} shows the codes that achieve lowest $\C$ for some values of $\dr$ for the scenario with an incoming process, where $\lambda_\text{c}=\mu=1$, $\omega/\mu=0.02$, $\rho=40$, and $\Gamma=3$. \ac{DS} is always beneficial, with replication and MDS codes performing the best for some $\Delta$, while regenerating and LRC codes perform poorer. Note that the discrepancies between the analytical and simulation results, in particular for $2$-replication, are due to the assumption in the computation of $\Cd$. However, the match is still very good.

\section{Conclusions}
\label{sec:conclusions}
We investigated the use of distributed storage in the mobile devices in a wireless network to reduce the communication cost of content delivery to the users. We introduced a repair scheduling where the repair of the data lost due to device departures is performed periodically. For this scenario, we derived analytical expressions for the overall communication cost, due to data download and repair, as a function of the repair interval. Using these expressions, we then investigated the performance of \ac{MDS} codes, regenerating codes and \acp{LRC}.

We showed that wireless \ac{DS} can reduce the overall communication cost with respect to the scenario where content is downloaded solely from the \ac{BS}. However, depending on the network parameters, there may exist a maximum value of the repair interval after which retrieving the file from the \ac{BS} is always less costly. Therefore, in such cases \ac{DS} is useful if repairs can be performed frequently enough. The required repair frequency depends on the network parameters and the code used for storage. In the case of an incoming process of nodes storing content, the communication cost using \ac{DS} can be further reduced. In this scenario, for large enough arrival rate of nodes bringing content, the use of wireless \ac{DS} with \ac{D2D} communication always entails a lower communication cost than downloading content only from the \ac{BS}. Interestingly, MDS codes yield better performance than codes specifically designed for \ac{DS}, such as regenerating codes and LRCs, if repair cannot be performed very frequently. The reason is that in this case a large tolerance to node failures and low repair locality is required. 

Our analysis shows that the use of erasure correcting codes to store (cache) content in the mobile devices is promising to reduce the communication cost of content delivery in a wireless network.


One interesting extension of this work is to consider the location of the mobile devices. In this case, the communication cost can be modeled as being dependent on the transmission distance. Another interesting extension of the analysis is to consider a library of files of varying popularity. For this scenario, one may analyze the use of different erasure correcting codes for files with different popularity, and exploiting multicast opportunities \cite{Ji2016}.

\appendices
\section{Proof of Theorem~\ref{thm:Cd}}
\label{prf:Cd}


To derive $\PrDd$ we first have to find the distribution of file requests within a repair interval $\dr$. Let $W_\ell$ be the time instant of the $\ell$th request and let $\Wlt\triangleq W_\ell\mod\dr$ be the time of the $\ell$th request in relation to a repair interval. The \ac{pdf} of $\Wlt$ is given by the following lemma.
\begin{lemma}
	\label{lem:Wlt}
	The distribution of $\Wlt$ for $t\in[0,\dr)$ is
	\begin{equation}
		\label{eq:Wltpdf}
		f_{\Wlt}(t) = \frac{\omega^\ell e^{-\omega t}}{(\ell-1)!} \sum_{i=0}^\infty (t+i\dr)^{\ell-1}e^{-\omega \dr i}.
	\end{equation}
\end{lemma}

\begin{IEEEproof}
	$W_\ell$ is computed as the sum of $\ell$ inter-request times with pdf given by \eqref{eq:Trpdf}. Thus, $W_\ell$ is an Erlang distributed random variable with pdf \cite[Sec.~3.4.5]{Miller2004}
	\begin{equation}\label{eq:Wlpdf}
		f_{W_\ell}(t)=\frac{\omega^\ell t^{\ell-1}e^{-\omega t}}{(\ell-1)!},\quad t\geq0.
	\end{equation}
	
	The transformation $g:W_\ell\to\Wlt$ is given by $t = g(x)$, where
	\begin{equation}
		\label{eq:g}
		g(x)=x-i\dr,\quad x\in[i\dr,(i+1)\dr),~i\ge0.
	\end{equation}
	Note that $g'(x)=1$ for $x\in(i\dr,(i+1)\dr)$. Moreover, $\lim_{x\to i\dr_-}g'(x)=\lim_{x\to i\dr_+}g'(x)=1$ and $g'(x)$ is continuous and well defined. Let $x_i$ be the roots of \eqref{eq:g},
	\begin{equation*}
		x_i=g^{-1}(t)=t+i\dr,\quad t\in[0,\dr).
	\end{equation*}
Then, \cite[Th. 4.2]{Miller2004}
	\begin{equation*}
		f_{\Wlt}(t) = \sum_{x_i}f_{W_\ell}(x_i)\left|\frac{1}{g'(x_i)}\right| = \sum_{i=0}^\infty f_{W_\ell}(t+i\dr),
	\end{equation*}
and \eqref{eq:Wltpdf} is obtained using \eqref{eq:Wlpdf}.
\end{IEEEproof}

Define $\Winf \triangleq \lim_{\ell\to\infty}\Wlt$. We have the following result.

\begin{lemma}
	\label{lem:Winf}
	The distribution of $\Winf$ for $t\in[0,\dr)$ is
	\begin{equation*}
		f_{\Winf}(t) = \frac{1}{\dr},
	\end{equation*}
	and the limit is achieved exponentially fast in $\ell$.
\end{lemma}

\begin{IEEEproof}
	Using Lerch's transcendent \cite[Sec.~25.14]{Olver2010}
	\begin{equation*}
		\Phi\left(e^{-\omega\dr},1-\ell,\frac{t}{\dr}\right)\triangleq\sum_{i=0}^\infty\left(\frac{t}{\dr}+i\right)^{\ell-1}e^{-\omega\dr i},\quad\ell>1,
	\end{equation*}
	the pdf of $\Wlt$ (Lemma~\ref{lem:Wlt}) can be rewritten as
	\begin{equation*}
		f_{\Wlt}(t) = \frac{(\omega\dr)^\ell e^{-\omega t}}{\dr\cdot(\ell-1)!}\Phi\left(e^{-\omega\dr},1-\ell,\frac{t}{\dr}\right).
	\end{equation*}

According to \cite[Cor.~4]{Navas2013},
	\begin{equation*}
		\lim_{\ell\to\infty}\frac{(\omega\dr)^\ell}{(\ell-1)!}\Phi\left(e^{-\omega\dr},1-\ell,\frac{t}{\dr}\right) = e^{\omega t}.
	\end{equation*}
	Hence, for an infinite number of requests
	\begin{equation*}
		\lim_{\ell\to\infty}f_{\Wlt}(t) = \frac{e^{-\omega t}}{\dr}\lim_{\ell\to\infty}\frac{(\omega\dr)^\ell}{(\ell-1)!}\Phi\left(e^{-\omega\dr},1-\ell,\frac{t}{\dr}\right)=\frac{1}{\dr}.
	\end{equation*}
	Furthermore, using \cite[Th. 3]{Navas2013}, as $\ell\rightarrow\infty$,
	\begin{equation}
		f_{\Wlt}(t)\le\frac{1}{\dr}+O\Bigg(\left(\frac{\sqrt{4\pi^2+(\omega\dr)^2}}{\omega\dr}\right)^{-\ell}\Bigg),
	\end{equation}
where $\frac{\sqrt{4\pi^2+(\omega\dr)^2}}{\omega\dr}\ge 1$. Therefore, the convergence is exponentially fast in $\ell$.
\end{IEEEproof}

%

We proceed with the second step of the proof. Within a repair interval, the number of storage nodes $m(t)$ in the cell is described by a Poisson death process \cite[Sec.~8.6]{Miller2004}. Denote by $T_i$ the time interval for which $m(t)=i$, $i = h, \ldots, m$ (see Fig. \ref{fig:stop} for an illustration). Note that $T_i$ is exponentially distributed with rate $\mu_i=i\mu$, since there are $i$ storage nodes in the cell and the departure rate per node is $\mu$ (see Section~\ref{sec:model}). Denote by $S_h$ the time instant at which $m(t)$ changes from $h$ to $h-1$, i.e., the time after which \ac{D2D} download is no longer possible.
$S_h$ can be written as
\begin{equation*}
	S_h = \sum_{i=h}^{m}T_i.
\end{equation*}
The pdf of $S_h$ is given by \cite[Sec.~1.3.1]{Bolch2006}
\begin{equation}
	\label{eq:Shpdf}
	f_{S_h}(t)=\sum_{i=h}^m\frac{\mu_m\mu_{m-1}\ldots\mu_{h}}{\prod_{\substack{j=h\\j\neq i}}^m(\mu_j-\mu_i)}e^{-\mu_it},\quad t\geq0.
\end{equation}
Note that $\Pr(S_h\ge\dr)>0$ for finite $\dr$, which implies that, with some probability, $m(t)\ge h$ for the duration of the repair interval. In this case, $\PrDd=1$.

We now have all the prerequisites to derive $\PrDd$. \ac{D2D} download is possible if at least $h$ storage nodes are available in the cell. Thus,
\begin{equation*}
	\PrDd = \lim_{L \rightarrow \infty}\frac{1}{L}\sum_{\ell=1}^{L}\Pr\left(\Wlt <S_h\right).
\end{equation*}
From the convergence result of Lemma~\ref{lem:Winf}, it follows that
\begin{align*}
	\PrDd & = \Pr\left(\Winf <S_h\right) = \Pr\left(\Winf-S_h<0\right)\\
		& = \int_{-\infty}^0 f_{\Winf-S_h}(t)~dt,
\end{align*}
where \cite{Miller2004}
\begin{equation*}
	f_{\Winf-S_h}(t) = \int_{-\infty}^{\infty}f_{\Winf}(t+s)f_{S_h}(s)~ds.
\end{equation*}

Using the results of Lemma~\ref{lem:Winf} and \eqref{eq:Shpdf}, we get after some calculation

\begin{align}
	\PrDd & = \frac{1}{\dr}\sum_{i=h}^m\int_{-\infty}^0e^{\mu_it}dt\left(1-e^{-\mu_i\dr}\right)\prod_{\substack{j=h\\j\neq i}}^m\frac{j}{j-i}\nonumber\\
	\label{eq:PrDd}
	& = \frac{1}{\dr}\sum_{i=h}^m\frac{1-p_i}{\mu_i}\prod_{\substack{j=h\\j\neq i}}^m\frac{j}{j-i}.
\end{align}

By inserting \eqref{eq:PrDd} into \eqref{eq:Cda} and using $\PrDd+\PrBSd=1$, we obtain \eqref{eq:Cd}.


\section{Proof of Corollary \ref{cor:Casym}}
\label{prf:Casym}
Consider the case $\dr\to0$. For the repair cost (Theorem~\ref{thm:Cr}),
\begin{align*}
	\lim_{\dr\to0}\Cr & = \frac{1}{F}\left(\rhoBS \gBS \sum_{i=0}^{r-1}(m-i) \lim_{\dr\to0}\frac{b_i(m, p)}{\dr}\right.\\
		& \left.+\rhoD\gD \sum_{i=r}^m(m-i) \lim_{\dr\to0}\frac{b_i(m, p)}{\dr}\right),
\end{align*}
where $b_i(m,p)$ is given in \eqref{eq:binom} and $p=e^{-\mu \dr}$. Note that
\begin{align*}
	& \lim_{\dr\to0}\frac{b_i(m,p)}{\dr}\\
	& = \binom{m}{i}\lim_{\dr\to0}\frac{e^{-\mu\dr i}(1-e^{-\mu\dr})^{m-i}}{\dr}\\
	& \stackrel{(a)}{=} \mu\binom{m}{i}\lim_{\dr\to0}e^{-\mu\dr i}\left(1-e^{-\mu\dr}\right)^{m-i-1}\left(me^{-\mu\dr}-i\right)\\
	& =\left\{
		\begin{array}{ll}
			m\mu,\quad & \text{if}~i=m-1,\\
			0, & \text{otherwise}.
		\end{array}
	\right. ,
\end{align*}
where in $(a)$ we used l'H\^opital's rule.
Hence,
\begin{equation*}
	\sum_{i=0}^{r-1}(m-i)\lim_{\dr\to0}\frac{b_i(m,p)}{\dr}=0,
\end{equation*}
and
\begin{equation*}
	\sum_{i=r}^{m}(m-i)\lim_{\dr\to0}\frac{b_i(m,p)}{\dr}=(m-(m-1))m\mu=m\mu.
\end{equation*}
This implies
\begin{equation}\label{eq:Crinst}
	\lim_{\dr\to0}\Cr=\rhoD\gD m\mu.
\end{equation}

For the download cost (Theorem~\ref{thm:Cd}),
\begin{align}
	& \lim_{\dr\to0}\Cd = M\omega\Bigg(\rhoBS \nonumber\\
	& +\left(\rhoD\frac{h\alpha}{F}-\rhoBS\right)\sum_{i=h}^m\frac{1}{\mu_i}\lim_{\dr\to0}\frac{1-p_i}{\dr}\prod_{\substack{j=h\\j\neq i}}^m\frac{j}{j-i}\Bigg) \nonumber\\
	\label{eq:Cdinst}
	& = M\omega\Bigg(\rhoBS+\left(\rhoD\frac{h\alpha}{F}-\rhoBS\right)\sum_{i=h}^m\prod_{\substack{j=h\\j\neq i}}^m\frac{j}{j-i}\Bigg).
\end{align}
To simplify the expression, consider the function
\begin{equation}
	\label{eq:F}
	f(x)=\frac{1}{\prod_{i=h}^m(i-x)},
\end{equation}
which can be expanded as the sum of partial fractions as \cite[Ch. 6]{Adams2010}
\begin{equation}
\label{eq:Fb}
	f(x)=\sum_{i=h}^m\frac{1}{(i-x)\prod_{\substack{j=h\\j\neq i}}^m(j-i)}.
\end{equation}
Now, note that the sum in \eqref{eq:Cdinst} can be expressed as
\begin{equation*}
	\sum_{i=h}^m\prod_{\substack{j=h\\j\neq i}}^m\frac{j}{j-i} = \sum_{i=h}^m\frac{\prod_{j=h}^m j}{i\prod_{\substack{j=h\\j\neq i}}^m(j-i)} \stackrel{(a)}{=} f(0)\prod_{j=h}^m j \stackrel{(b)}{=} 1, 
\end{equation*}
where in $(a)$ we used \eqref{eq:Fb}, and in $(b)$ we used \eqref{eq:F}. Using this in \eqref{eq:Cdinst} we obtain
\begin{equation}
	\label{eq:Cdinst2}
	\lim_{\dr\to0}\Cd = M\omega\rhoD\frac{h\alpha}{F}.
\end{equation}
Finally, the expression \eqref{eq:Cinst} is obtained by using
\begin{equation*}
	\lim_{\dr\to0}\C = \lim_{\dr\to0}\Cr+\lim_{\dr\to0}\Cd.
\end{equation*}

%
	
Now, assume $\dr\to\infty$. For the average repair cost (Theorem~\ref{thm:Cr})
\begin{align*}
	\lim_{\dr\to\infty}\Cr & = \frac{1}{F}\left(\rhoBS \gBS\sum_{i=0}^{r-1}(m-i)\lim_{\dr\to\infty}\frac{b_i(m,p)}{\dr}\right.\\
		& ~~~~~~\left.+\rhoD \gD\sum_{i=r}^m(m-i)\lim_{\dr\to\infty}\frac{b_i(m,p)}{\dr}\right).
\end{align*}
Now,
\begin{equation*}
	\lim_{\dr\to\infty}\frac{b_i(m,p)}{\dr}=\binom{m}{i}\lim_{\dr\to\infty}\frac{e^{-\mu\dr i}(1-e^{-\mu\dr})^{m-i}}{\dr}=0,
\end{equation*}
which implies $\lim_{\dr\to\infty}\Cr=0$.

For the average download cost (Theorem~\ref{thm:Cd}),
\begin{align*}
	\lim_{\dr\to\infty}\Cd & = M\omega\Bigg[\rhoBS + \left(\rhoD\frac{h\alpha}{F}-\rhoBS\right)\\
	& ~~~~~~~~~~~\times \sum_{i=h}^m\frac{1}{\mu_i}\lim_{\dr\to\infty}\frac{1-p_i}{\dr}\prod_{\substack{j=h\\j\neq i}}^m\frac{j}{j-i}\Bigg],
\end{align*}
where $\mu_i = i\mu$, $p_i = e^{-\mu_i\dr}$. As $\lim_{\dr\to\infty}\frac{1-p_i}{\dr}=0~\forall~i$, then
\begin{equation*}
	\lim_{\dr\to\infty}\Cd = M\omega\rhoBS,
\end{equation*}
and \eqref{eq:Cno} follows.

\section{Proof of Theorem \ref{lem:Cdhbr}}
\label{prf:hbrdCd}
Following the proof of Theorem~\ref{thm:Cd} (Appendix~\ref{prf:Cd}), the probability that there are $m(t) = i$ storage nodes available at the time of a request is
\begin{align}
\label{eq:ci}
	\cc_i & \triangleq \Pr(S_{i+1}<\Winf<S_i) \nonumber\\
	& = \Pr(\Winf-S_i<0)-\Pr(\Winf-S_{i+1}<0).
\end{align}
The two probabilities in \eqref{eq:ci} can be obtained by replacing $h$ with $i$ and $i+1$ in \eqref{eq:PrDd},
\begin{align*}
	 \Pr(\Winf-S_i<0) &= \frac{1}{\dr}\sum_{i'=i}^m\frac{1-p_{i'}}{\mu_{i'}}\prod_{\substack{j=i\\j\neq i'}}^m\frac{j}{j-i'}\\
	\Pr(\Winf-S_{i+1}<0) &= \frac{1}{\dr}\sum_{i'=i+1}^m \frac{1-p_{i'}}{\mu_{i'}}\prod_{\substack{j=i+1\\j\neq i'}}^m \frac{j}{j-i'}.
\end{align*}
If no storage nodes are available, we always have to rely on \ac{BS} download. By replacing $h$ with 1 in \eqref{eq:PrDd}, we get that this occurs with probability
\begin{equation}
	\label{eq:PrBS}
	\PrBSd = 1-\frac{1}{\dr}\sum_{i=1}^m \frac{1-p_i}{\mu_i} \prod_{\substack{j=1\\j\neq i}}^m \frac{j}{j-i}.
\end{equation}
If $m(t)\ge h$, \ac{D2D} download is performed. This occurs with probability $\PrDd$, derived in Theorem~\ref{thm:Cd}.

For $m(t) = i, 1\le i\le h-1$, the hybrid scheme will achieve a lower download cost if $\rhoBS F > (\rhoBS(h-i)+i\rhoD)\alpha$, i.e., if
\begin{equation*}
	i > \frac{\rhoBS}{\rhoBS-\rhoD}\left(h-\frac{F}{\alpha}\right)\triangleq \dd.
\end{equation*}
Let
\begin{equation*}
	a \triangleq \min\left\{\left\lfloor \dd \right\rfloor, h-1\right\}.
\end{equation*}
For $1\le i\le a$, downloading $F$ bits from the \ac{BS} will give the lowest possible cost. For $a+1\le i\le h-1$, downloading $i\alpha$ bits through \ac{D2D} communication and $(h-i)\alpha$ bits from the \ac{BS} will give the lowest possible cost. The average download cost in the hybrid regime is hence
\begin{align}
\label{eq:Cdhp}
	\Cdh & = \frac{M\omega}{F}\left(\rhoBS F\PrBSd + \rhoBS F\sum_{i=1}^a \cc_i\right.\nonumber\\
		& \left.+ \sum_{i=a+1}^{h-1}(\rhoBS(h-i)+i\rhoD)\alpha\cc_i + \rhoD h\alpha\PrDd\right).
\end{align}
Finally, \eqref{eq:Cdhbr} is obtained by using \eqref{eq:PrDd} and \eqref{eq:PrBS} in \eqref{eq:Cdhp}.

\section{Proof of Theorem \ref{thm:LRCrep}}
\label{prf:LRCrep}

Recall that a storage node can be repaired {\itshape locally} or {\itshape globally} in \ac{D2D} communication. Only single node departures (within a repair group) can be repaired locally. Using \eqref{eq:binom}, the average number of local \ac{D2D} repairs in a repair group is
\begin{equation*}
	b_r(r+1,p) = (r+1)p^r(1-p),
\end{equation*}
where $p=e^{-\mu\dr}$. Since there are $G = \frac{m}{r+1}$ disjoint repair groups, the average number of local \ac{D2D} repairs per $m$ storage nodes is
\begin{equation*}
	\mrlD = G(r+1)p^r(1-p) = mp^r(1-p).
\end{equation*}
This entails a cost $\rhoD\gD\mrlD$ [c.u.].

We now compute the average number of global \ac{D2D} repairs $\mrgD$. Let $\X=(X_0, X_1, \ldots, X_{r+1})$, where $X_i\in\{0,1,\ldots,G\}$, $\sum_iX_i=G$, is the random variable giving the number of repair groups with $i$ storage node departures in a repair interval $\dr$. The number of global repairs is given by $\sum_{i=2}^{r+1}i X_i$, under the constraint that there are at least $h$ storage nodes available at the time of a repair, i.e., if $\sum_{i=1}^{r+1} iX_i\le m-h$. Therefore, by averaging over all possible realizations $\x = (x_0, x_1, \ldots, x_{r+1})$ of $\X$, we obtain
\begin{equation*}
	\mrgD = \sum_{\x:|\x|=G}\binom{G}{\x}\y^{\x} \cdot \sum_{i=2}^{r+1}ix_i \cdot \mathds{1}\left\{\sum_{i=1}^{r+1} ix_i\le m-h\right\},
\end{equation*}
where $|\x| \triangleq \sum_i x_i$, $\binom{G}{\x} \triangleq \frac{G!}{x_0!x_1!\cdots x_{r+1}!}$, and $\y^{\x} \triangleq \prod_i y_i^{x_i}$. The communication cost associated to global \ac{D2D} repairs is $\rhoD h\aLRC\mrgD$ [c.u.].

Finally, using \eqref{eq:binom}, the average total number of storage node departures in a repair interval is
\begin{equation*}
	\sum_{i=0}^m (m-i)b_i(m,p) = m(1-p).
\end{equation*}
All storage nodes that are not repaired in \ac{D2D} are repaired by the \ac{BS}. Therefore,
\begin{equation*}
	\mrBS = m(1-p)-\mrlD-\mrgD,
\end{equation*}
with communication cost $\rhoBS\gBS\mrBS$ [c.u.].

Finally, adding the three contributions $\rhoD\gD\mrlD$, $\rhoD h\aLRC\mrgD$ and $\rhoBS\gBS\mrBS$, and dividing by $\dr$ and normalizing by $F$, we obtain \eqref{eq:CrLRC}.

\section{Proof of Lemma \ref{lem:optcode}}
\label{prf:optcode}

The overall communication cost for $\dr=0$ is (Corollary~\ref{cor:Casym})
\begin{equation}
	\label{eq:limC}
	\lim_{\dr\to0}\C = \frac{\rhoD}{F}(\gD m\mu+M\omega h\alpha).
\end{equation}

Consider an $[m,h,r]$ linear code with $m=n$ and minimum Hamming distance $d\ge 2$. It follows that $\alpha=\frac{F}{k}$, $\beta=\alpha$, and $h\ge k$, where the equality is achieved for MDS codes. Furthermore, note that $d=m-h+1$. Also, from \cite{Gopalan2012},
\begin{equation}
	\label{eq:Gop}
	d\le n-k-\left\lceil\frac{k}{r}\right\rceil+2.
\end{equation}
Using $m=n$ and the fact that $d\ge 2$ in \eqref{eq:Gop}, we can write
\begin{equation*}
	m\ge k+\left\lceil\frac{k}{r}\right\rceil\ge k+\frac{k}{r}.
\end{equation*}
Now, using this, $\gD=r\beta=r\alpha$, and $\alpha=\frac{F}{k}$ in \eqref{eq:limC} we obtain
\begin{align}
\label{eq:BoundC}
	\lim_{\dr\to0}\C &= \frac{\rhoD}{F}(\gD m\mu+M\omega h\alpha)\nonumber\\
	&=\rhoD\left(\frac{r}{k} m\mu+M\omega \frac{h}{k}\right)\nonumber\\
	&\ge \rhoD\left((r+1)\mu+M\omega \frac{h}{k}\right)\nonumber\\
	&\ge \rhoD\left(2\mu+M\omega \right),
\end{align}
where in the last inequality we used $r\ge 1$ and $h\ge k$. It is easy to verify that the lower bound in \eqref{eq:BoundC} is achieved by $2$-replication.

Now, consider \acp{LRC}. We get
\begin{equation*}
	m\gD = F\frac{m}{h}(r+1) > 2F,
\end{equation*}
since $h<m$ and $r\ge1$. Also,
\begin{equation*}
	h\aLRC = F\frac{r+1}{r} > F.
\end{equation*}
Inserting this into \eqref{eq:limC} gives that \acp{LRC} yield a higher overall communication cost than \eqref{eq:BoundC}.

Consider now \ac{MBR} codes. We would like to minimize $m\gD$ under the constraints $m\ge2$, $h\ge1$, and $h<m$, for $r=m-1$. For $h=m-1$, $m\gD=2F$. For $h<m-1$, relaxing the integer constraints on $m$ and $h$,
\begin{equation*}
	\frac{\partial}{\partial m}~m\gD = 4\frac{F}{h}\frac{m^2-m(h+1)+1}{(2m-h-1)^2}>0.
\end{equation*}
Consequently, $m\gD$ is minimized for $h = m-1$ and the minimum is equal to $2F$. We proceed to minimize $h\aMBR$ for $r=m-1$ under the same constraints. For $h=1$, we have $h\aMBR=F$. Also, for $h>1$,
\begin{equation*}
	\frac{\partial}{\partial h}~h\aMBR = 2F\frac{m-1}{(2m-h-1)^2}>0.
\end{equation*}
As a result, $m\gD$ and $h\aMBR$ are jointly minimized for $m=2$ and $h=1$. Thus, the \ac{MBR} code, which is indeed $2$-replication, achieves the lower bound in \eqref{eq:BoundC}.

We proceed to investigate the overall communication cost when $\dr=0$ for \ac{MSR} codes. By setting $r=m-1$ we minimize $\gD$ with respect to $r$. We relax the integer constraints on $m$ and $h$. By differentiating $m\gD$ with respect to $h$ and setting the derivative equal to zero, we find
\begin{equation*}
	\arg\min_{h}~m\gD = \frac{m}{2}.
\end{equation*}
Under the constraints $m\ge2$, $h\ge1$ and $h<m$, we have
\begin{equation*}
	\left.\frac{\partial}{\partial m}~m\gD\right|_{m=2h} = \frac{F}{h^2} > 0.
\end{equation*}
This implies that $m\gD$ is minimized for $m=2$ and $h=1$ and that the minimum is equal to $2F$. Since $h\aMSR=F$, $m\gD$ and $h\aMSR$ are jointly minimized for $m=2$ and $h=1$. Therefore, the $[2,1,1]$ \ac{MSR} code, which corresponds to $2$-replication, achieves the lower bound in \eqref{eq:BoundC}. This concludes the proof.

\balance
\bibliographystyle{IEEEtran}

\end{document}